\documentclass{emulateapj}
\usepackage{psfig,enumerate}
\usepackage{url}
\usepackage{verbatim}
\usepackage{color}
\usepackage{lscape}
\usepackage{atbegshi}
\def\hackaltaffiltext#1#2{\AtBeginShipoutNext{\footnotetext[#1]{#2}\stepcounter{footnote}}}

\newcommand{\dgr}{$^{\circ}~$}

\newcommand{\hi}{H{\footnotesize I} }

\begin{document}

\title{SMASH -- Survey of the MAgellanic Stellar History}

\shorttitle{SMASH Overview}
\shortauthors{Nidever et al.}

\author{
David L. Nidever\altaffilmark{1,2,3,4},
Knut Olsen\altaffilmark{1},
Alistair R. Walker\altaffilmark{5},
A. Katherina Vivas\altaffilmark{5},
Robert D. Blum\altaffilmark{1},
Catherine Kaleida\altaffilmark{6},
Yumi Choi\altaffilmark{3},
Blair C. Conn\altaffilmark{7,8},
Robert A. Gruendl\altaffilmark{9,10},
Eric F. Bell\altaffilmark{4},
Gurtina Besla\altaffilmark{3},
Ricardo R. Mu\~noz\altaffilmark{11,12},
Carme Gallart\altaffilmark{13,14},
Nicolas F. Martin\altaffilmark{15,16},
Edward W. Olszewski\altaffilmark{3},
Abhijit Saha\altaffilmark{1},
Antonela Monachesi\altaffilmark{17,18,19},
Matteo Monelli\altaffilmark{13,14},
Thomas J. L. de Boer\altaffilmark{20},
L. Clifton Johnson\altaffilmark{21},
Dennis Zaritsky\altaffilmark{3},
Guy S. Stringfellow\altaffilmark{22},
Roeland P. van der Marel\altaffilmark{6},
Maria-Rosa L. Cioni\altaffilmark{23,24,25},
Shoko Jin\altaffilmark{26},
Steven R. Majewski\altaffilmark{27},
David Martinez-Delgado\altaffilmark{28},
Lara Monteagudo\altaffilmark{13,14},
Noelia E. D. No\"el\altaffilmark{29},
Edouard J. Bernard\altaffilmark{30},
Andrea Kunder\altaffilmark{24},
You-Hua Chu\altaffilmark{31,10},
Cameron P. M. Bell\altaffilmark{24},
Felipe Santana\altaffilmark{11},
Joshua Frechem\altaffilmark{32},
Gustavo E. Medina\altaffilmark{11},
Vaishali Parkash\altaffilmark{33},
J. C. Ser\'on Navarrete\altaffilmark{5},
Christian Hayes\altaffilmark{27}
}

\altaffiltext{1}{National Optical Astronomy Observatory, 950 North Cherry Ave, Tucson, AZ 85719 (dnidever@noao.edu)}
\altaffiltext{2}{Large Synoptic Survey Telescope, 950 North Cherry Ave, Tucson, AZ 85719}
\altaffiltext{3}{Steward Observatory, University of Arizona, 933 North Cherry Avenue, Tucson AZ, 85721}
\altaffiltext{4}{Department of Astronomy, University of Michigan, 1085 S. University Ave., Ann Arbor, MI 48109-1107, USA}
\altaffiltext{5}{Cerro Tololo Inter-American Observatory, National Optical Astronomy Observatory, Casilla 603, La Serena, Chile}
\altaffiltext{6}{Space Telescope Science Institute, 3700 San Martin Drive, Baltimore, MD 21218}
\altaffiltext{7}{Research School of Astronomy and Astrophysics, Australian National University, Canberra, ACT 2611, Australia}
\altaffiltext{8}{Gemini Observatory, Recinto AURA, Colina El Pino s/n, La Serena, Chile.}
\altaffiltext{9}{National Center for Supercomputing Applications, 1205 West Clark St., Urbana, IL 61801, USA}
\altaffiltext{10}{Department of Astronomy, University of Illinois, 1002 West Green St., Urbana, IL 61801, USA}
\altaffiltext{11}{Departamento de Astronom\'ia, Universidad de Chile, Camino del Observatorio 1515, Las Condes, Santiago, Chile}
\altaffiltext{12}{Visiting astronomer, Cerro Tololo Inter-American Observatory, National Optical Astronomy Observatory, which is operated by the Association of Universities for Research in Astronomy (AURA) under a cooperative agreement with the National Science Foundation.}
\altaffiltext{13}{Instituto de Astrof\'{i}sica de Canarias, La Laguna, Tenerife, Spain}
\altaffiltext{14}{Departamento de Astrof\'{i}sica, Universidad de La Laguna, Tenerife, Spain}
\altaffiltext{15}{Universit\'e de Strasbourg, CNRS, Observatoire astronomique de Strasbourg, UMR 7550, F-67000 Strasbourg, France}
\hackaltaffiltext{16}{Max-Planck-Institut f\"ur Astronomie, K\"onigstuhl 17, D-69117 Heidelberg, Germany}
\hackaltaffiltext{17}{Max-Planck-Institut f\"ur Astrophysik, Karl-Schwarzschild-Str. 1, 85748 Garching, Germany}
\hackaltaffiltext{18}{Instituto de Investigaci\'on Multidisciplinario en Ciencia y Tecnolog\'ia, Universidad de La Serena, Ra\'ul Bitr\'an 1305, La Serena, Chile}
\hackaltaffiltext{19}{Departamento de F\'isica y Astronom\'ia, Universidad de La Serena, Av. Juan Cisternas 1200 N, La Serena, Chile}
\hackaltaffiltext{20}{Institute of Astronomy, University of Cambridge, Madingley Road, Cambridge CB3 0HA, UK}
\hackaltaffiltext{21}{Center for Astrophysics and Space Sciences, UC San Diego, 9500 Gilman Drive, La Jolla, CA, 92093-0424, USA}
\hackaltaffiltext{22}{Center for Astrophysics and Space Astronomy, University of Colorado, 389 UCB, Boulder, CO, 80309-0389, USA}
\hackaltaffiltext{23}{Universit\"{a}t Potsdam, Institut f\"{u}r Physik und Astronomie, Karl-Liebknecht-Str. 24/25, 14476 Potsdam, Germany}
\hackaltaffiltext{24}{Leibniz-Institut f\"{u}r Astrophysics Potsdam (AIP), An der Sternwarte 16, 14482 Potsdam Germany}
\hackaltaffiltext{25}{University of Hertfordshire, Physics Astronomy and Mathematics, Hatfield AL10 9AB, United Kingdom}
\hackaltaffiltext{26}{STFC RALSpace, Rutherford Appleton Laboratory, Harrell Oxford OX11 OQX UK}
\hackaltaffiltext{27}{Department of Astronomy, University of Virginia, Charlottesville, VA 22904, USA}
\hackaltaffiltext{28}{Astronomisches Rechen-Institut, Zentrum f\"ur Astronomie der Universit\"at Heidelberg,  M{\"o}nchhofstr. 12-14, 69120 Heidelberg, Germany}
\hackaltaffiltext{29}{Department of Physics, University of Surrey, Guildford, GU2 7XH, UK}
\hackaltaffiltext{30}{Universit\'e C\^ote d'Azur, OCA, CNRS, Lagrange, France}
\hackaltaffiltext{31}{Institute of Astronomy and Astrophysics, Academia Sinica, No.1, Sec. 4, Roosevelt Rd, Taipei 10617, Taiwan, R.O.C.}
\hackaltaffiltext{32}{Rochester Institute of Technology, Rochester, NY 14623}
\hackaltaffiltext{33}{Monash Centre for Astrophysics, School of Physics and Astronomy, Monash University, Victoria 3800, Australia}


\begin{abstract}
The Large and Small Magellanic Clouds are unique local laboratories for studying the formation and evolution of small galaxies in exquisite detail.
The Survey of the MAgellanic Stellar History (SMASH) is an NOAO community DECam survey of the Clouds mapping 480 deg$^2$ (distributed
over $\sim$2400 square degrees at $\sim$20\% filling factor) to ~24th mag in $ugriz$.  The primary goals of SMASH are to identify low surface
brightness stellar populations associated with the stellar halos and tidal debris of the Clouds, and derive spatially-resolved star formation histories.
Here, we present a summary of the survey, its data reduction, and a description of the first public Data Release (DR1).  The SMASH DECam data
have been reduced with a combination of the NOAO Community Pipeline, the PHOTRED automated PSF photometry pipeline, and custom calibration
software.  The astrometric precision is $\sim$15 mas and the accuracy is $\sim$2 mas with respect to the Gaia reference frame. The photometric
precision is $\sim$0.5--0.7\% in $griz$ and $\sim$1\% in $u$ with a calibration accuracy of $\sim$1.3\% in all bands.  The median 5$\sigma$ point
source depths in $ugriz$ are 23.9, 24.8, 24.5, 24.2, 23.5 mag.  The SMASH data already have been used to discover the Hydra II Milky Way satellite,
the SMASH 1 old globular cluster likely associated with the LMC, and extended stellar populations around the LMC out to R$\sim$18.4 kpc.
SMASH DR1 contains measurements of $\sim$100 million objects distributed in 61 fields.   A prototype version of the NOAO Data Lab provides
data access and exploration tools.
\end{abstract}


\section{Introduction}
\label{sec:intro}

The Large and Small Magellanic Clouds (LMC and SMC), as two of the nearest and most massive satellite galaxies of the Milky Way (MW), offer a unique opportunity
to study the processes of galaxy formation and evolution of low-mass galaxies in great detail.
The Clouds have long held broad importance for astronomy, both as laboratories of astrophysical processes and as calibrators of the extragalactic distance scale. 
As the closest example of an
interacting pair of galaxies, they provide special insight into the impact of such interactions on the structure and evolution of galaxies.

The Clouds are ideally suited to addressing some particularly critical questions:
What are the consequences of stripping of stars and gas when dwarf galaxies fall into the halos of more massive galaxies, an important mode of mass growth
for galaxies since $z$$\sim$1?  What are the properties of the hot and warm gaseous halos of galaxies like the Milky Way, the density of which sets the efficiency
of gas stripping and ``quenching" of satellites?  What are the physical mechanisms and timescales, if any, behind the triggering of star formation by galaxy interactions?  These questions can only be addressed by surveys that probe the stars, gas, and dust of the Magellanic Clouds and their surroundings.

\citet{deV55} began the systematic study of the Magellanic Clouds (MCs),
highlighting the young and bright structures that trace the LMC's bar and the SMC's very irregular shape.
Although the gaseous component of the LMC's disk extends only to a radius of $\sim$4\dgr \citep{SS03}, the stellar component stretches over a much larger area.
Stellar catalogs from the large near-infrared surveys DENIS \citep{Epchtein97} and 2MASS \citep{Skrutskie06} show it extending to at least $\sim$8\dgr and 
were used to measure the LMC disk structure and viewing angle \citep{vdM01,vdMCioni01}.  Furthermore, carbon stars from \citet{Kunkel97} follow disk kinematics
out to $\sim$13\dgr \citep{vdM02}.  Finally, the Outer Limits Survey \citep[OLS;][]{Saha10} used deep photometry to study old LMC main-sequence stars in select fields far from the LMC center,
and found that in the north they followed the disk exponential profile out to $\sim$16\degr.

Indeed, the periphery of the Magellanic Clouds have been a rich ground for new discoveries.  In the LMC,
\citet{Minniti03} found kinematical evidence for an old stellar halo using RR Lyrae stars (in the central region) with a large velocity dispersion.
In the LMC outskirts, \citet{Munoz06} discovered a kinematically cold group of LMC stars in the foreground of the Carina dwarf spheroidal galaxy.
Subsequently, the MAgellanic Periphery Survey (MAPS; described in \citealt{Nidever11}) used spectroscopically-confirmed red giant branch stars to detect
a halo-like stellar population extending out to $\sim$22\dgr (over an $\sim$180\dgr azimuthal range) following a shallow de Vaucouleurs profile \citep{Majewski09}.
More recently, \citet{Mackey16} used the Dark Energy Survey \citep[DES;][]{DES} data to detect an ``arc-like" structure in the periphery of the LMC (at $\sim$15\dgr
from the center) which is likely a tidally disturbed portion of the LMC disk due to the recent interaction with the SMC.
In addition,  \citet{Belokurov16} used the DES data to map streams of Blue Horizontal Branch (BHB) stars around the MCs with some extending to $R$$\sim$40\degr.

In the SMC, early photometric plate photometry showed that the young blue stars (produced by recent intense star formation) have a very irregular
and patchy distribution while the older population (traced by giant stars) is fairly azimuthally symmetric and extends to larger radii
\citep[$\sim$5\degr;][]{HH89,GH91,GH92}, as expected for dIrr galaxies \citep{Mateo98}.
Subsequent deep CCD photometry by \citet{NG07} found old stellar populations out 5.8\dgr
and \citet{DePropris10} detected red giant stars spectroscopically out to $\sim$6\degr.  The MAPS survey used photometrically selected
red giant stars to find that the older SMC stellar population followed the same exponential profile from the center of the SMC out to
$\sim$7.5\dgr but that beyond that there was a break in the radial profile with stars extending out to $\sim$11\dgr (later confirmed with DES data by
\citealt{Pieres17}) potentially representing a classical stellar halo of the SMC.
In addition, \citet{Nidever13} found evidence for a stellar component of the tidally stripped Magellanic Bridge and, more recently, 
\citet{BK16} used Gaia DR1 data to find both young and old stellar bridges, though not spatially coincident, between the Clouds.
Finally, \citet{Pieres17} used DES data to discover a stellar overdensity in the northern region of the SMC (at $\sim$8\degr) that might be a tidally stripped dwarf
galaxy.

Explanations for the structure found in the periphery of the Magellanic Clouds must naturally account for the complex history of interaction that they have had with each other and the Milky Way.  These interactions, which have produced the gaseous Magellanic Stream, Leading Arm and Bridge
\citep{Putman03, Stani99, Muller03, Br05, Nidever08, Nidever10}, have been explored for over three decades \citep[e.g.,][]{MF80,GN96,YN03,Connors04,Connors06,Mastro05,Besla12,Besla13}.
Many of the features of the Magellanic system were well reproduced by a model invoking tidal stripping through repeated close passages to the
MW by the MCs on their bound orbit \citep[e.g.,][]{GN96}, although ram pressure stripping \citep[e.g.,][]{Mastro05}
and star formation feedback \citep[e.g.,][]{Olano04,Nidever08} were also put forward as origin mechanisms and are likely important factors.
However, recent HST-based proper motion measurements of the MCs \citep{Kallivayalil06a,Kallivayalil06b,Kallivayalil13} indicate that the MCs have higher velocities relative to the
Milky Way than previously thought.  Such high velocities may allow for multiple interactions of the Clouds with the Milky Way \citep{Bekki11, Gomez15}.  A more likely scenario, however, is that the MCs are approaching the MW environment for the first time \citep{Besla07,Gomez15}, as favored by cosmological simulations \citep{Boylan-Kolchin11,Busha11,Patel17}. This discovery has forced a reinterpretation of many features of the Magellanic System, leading recent simulations \citep{Besla10, Besla12, DB12} to suggest that LMC-SMC interactions alone are responsible for the formation of the Magellanic Stream, including a potential direct collision between the LMC and SMC. The discovery that the LMC has stripped a large number of stars from the SMC \citep[$\sim$5\% of the LMC's mass,][]{Olsen11} supports the collision hypothesis, or at least necessitates a very close interaction between the SMC and LMC. If that collision also produced the Stream and Leading Arm, which are now known to extend for at least 200\dgr across the sky \citep{Nidever10}, then there ought to be stellar tidal debris over a similar area of sky as well, with potentially high densities in the area of the Leading Arm \citep{Besla10}.

The primary aim of the Survey of the MAgellanic Stellar History (SMASH) is to trace and to measure the extended stellar populations of the Clouds, including potential tidal debris, extended disks, and halo components.   SMASH is an NOAO survey project that is using the Dark Energy Camera \citep[DECam;][]{Flaugher15} on the NOAO Blanco 4$-$m telescope at Cerro Tololo Interamerican Observatory (CTIO) to observe a total area of 
480 deg$^2$ of the Magellanic periphery and the main bodies with deep $ugriz$ images.  SMASH fields completely cover the main bodies of the Clouds, while in the periphery they are distributed as ``islands'' over an area of $\sim$2400 deg$^2$, to maximize the chances of detecting widely distributed stellar populations within the allotted observing time.  SMASH builds on the technique first adopted by the Outer Limits Survey \citep{Saha10}, using old main sequence stars as tracers of populations down to surface brightnesses equivalent to $\Sigma_g$=35 mag arcsec$^{-2}$.  Compared to the $\sim$5000 deg$^2$ Dark Energy Survey (DES), SMASH covers an order of magnitude less total area, but probes a region that is only a factor of two smaller in size.  Whereas DES covers the northern MC periphery and includes the Magellanic Stream, SMASH covers the southern periphery (as well as the MC main bodies) and includes the Leading Arm, where models \citet{Besla10} predict that tidal debris is prominent.  SMASH photometry is as deep or deeper than DES in $griz$, and includes the $u$ filter, which DES does not, but does not include the $Y$ filter.

The SMASH survey addresses a number of important scientific goals, for which progress is underway.  The detection of stellar debris in either the Leading Arm or Magellanic Stream would make them the only tidal streams
with known gaseous and stellar components in the Local Group.  This would not only be invaluable for understanding the history and observable consequences of the Magellanic interaction, but would give us a ``clean" dynamical tracer of the MW's dark halo and a way to prove the efficiency of the MW's hot halo gas to induce ram pressure effects.  The size of the LMC's stellar periphery is a direct probe of the tidal radius of the LMC, with which we can explore the dark matter halos of both the LMC and MW.  DES has led to the discovery of many new satellite galaxies
\citep{Bechtol15,Koposov15,Drlica-Wagner15}, some of which are likely associated with the Magellanic Clouds \citep{Deason15,Jethwa16,Walker16,Sales16}.  The discovery that satellite galaxies may have their own satellite systems is an important new piece to the puzzle of the ``missing satellites" \citep{Moore99,Klypin99}.  We have discovered one new satellite galaxy in SMASH, Hydra II \citep{Martin15},
which from its position and distance may be associated with the Leading Arm of the Magellanic Stream; the search for more is in progress.  Finally, it is known that strong population gradients exist to large radii in both MCs \citep{Gallart08, Cioni09}.  With SMASH, we are working to derive spatially resolved, precise star formation histories covering all ages of the MCs and to the largest radii, thus providing detailed information on their complete evolution.  We are also working to identify new star clusters and to map the MC's galactic structures.

\begin{figure*}[ht]
\includegraphics[width=1.0\hsize,angle=0]{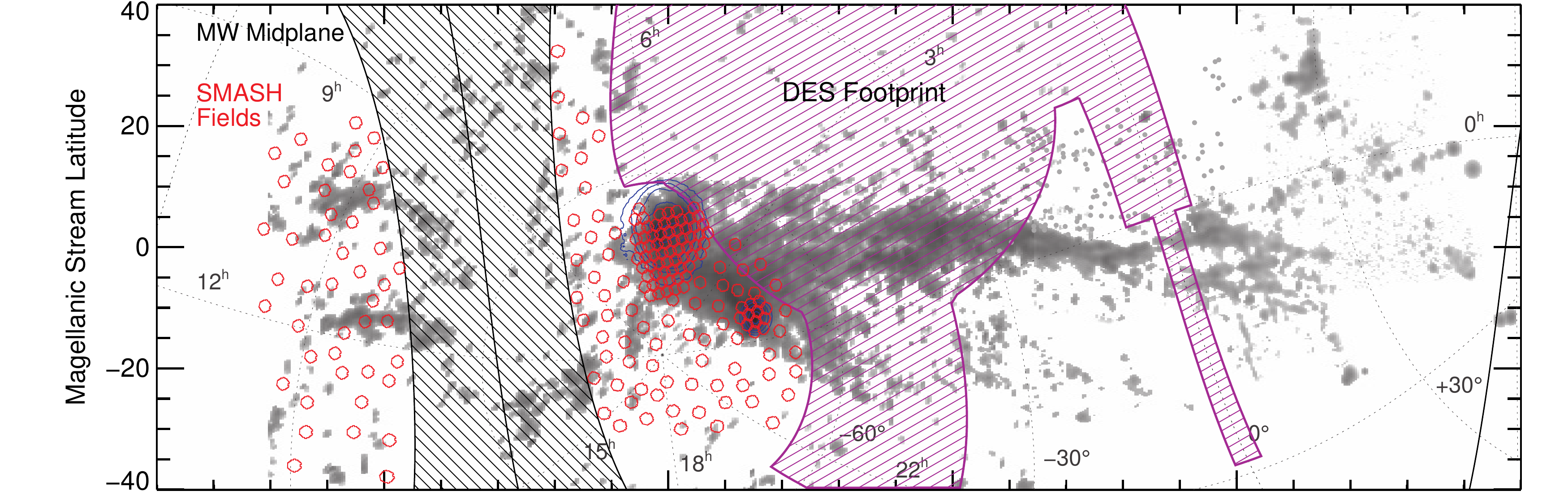}
\includegraphics[width=1.0\hsize,angle=0]{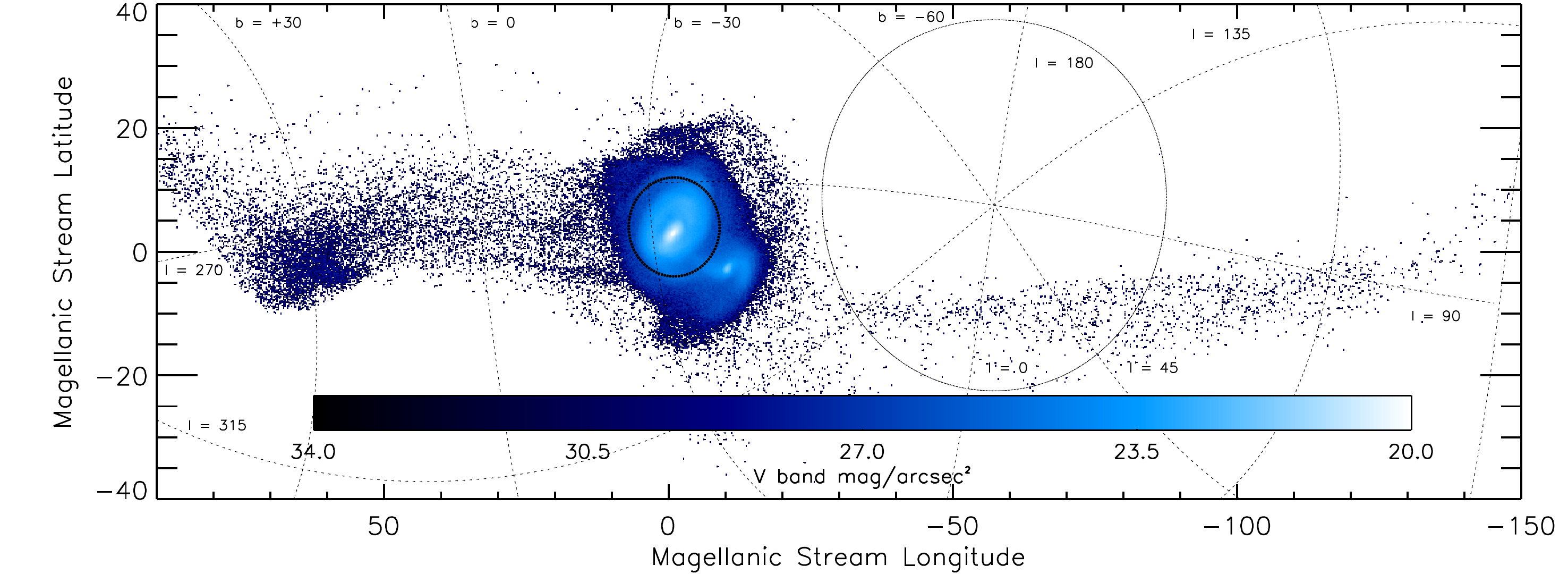}
\caption{\small
The region of the sky relevant to the Magellanic system in the Magellanic Stream coordinates system \citep{Nidever08}.
(Top) The observed \ion{H}{1} column density of the entire 200$^{\circ}$ Magellanic
Stream system \citep{Nidever10} is shown in grayscale, while the blue contours
represent the 2MASS \citep{Skrutskie06} red giant branch starcounts.
The open red hexagons are the SMASH survey fields (with the correct size and shape).
The DES footprint is represented by the purple shaded region.  The solid line represents the Galactic equator, the hashed region is a Galactic Zone of Avoidance region,
and the dotted lines are an equatorial coordinate grid.
(Bottom) The predicted $V$-band surface brightness (mag/arcsec$^2$) of the stellar component of the
Magellanic system from \citet{Besla13}.  The simulation predicts stellar structures out to large radii from the main bodies of
the Magellanic Clouds (varying on small scales), and a higher stellar density in the Leading Arm than in the trailing Stream.  In the absence of strong
gas drag, the stellar and gas components are expected to be coincident.  In this model the exact location of the debris is not tuned to match the
observations, but the relative surface brightness of the various components is a robust prediction. A Galactic coordinate grid is shown in dotted lines.
}
\label{fig_bigmap}
\end{figure*}

Besides SMASH, there are other optical and NIR surveys of the Magellanic Clouds.  The fourth phase of the Optical Gravitational Lensing Experiment \citep[OGLE-IV;][]{Udalski15} is contiguously covering
the main bodies of the Magellanic Clouds as well as the Bridge region between them with relatively shallow $VI$ ($\sim$21 mag) photometry but with many repeat observations
that make it useful for studying the distribution of Magellanic variable stars \citep{Soszynski16, Skowron14, J-D16, J-D17}.
The VISTA survey of the Magellanic Clouds system \citep[VMC;][]{Cioni11} is a near-infrared $YJK_s$ survey of the main bodies of the Magellanic Clouds that, among other things, is being
used to study the star formation history \citep{Rubele15}, and proper motion \citep{Cioni16} of the Clouds.  In addition, the STEP survey using the VST telescope is an optical $gi$-band
($\sim$23.5 mag) survey of the SMC and Bridge region with repeat observations \citep{Ripepi14}
As previously mentioned, the contiguous DES survey serendipitously covers the northern
regions of the Magellanic Clouds as well as the Magellanic Stream and has been used to study the Magellanic system \citep{BK16, Mackey16}.
Finally, the Magellanic Satellites Survey \citep[MagLiteS; described in ][]{Drlica-Wanger16} is searching for Magellanic satellites using shallow $g$ and $r$ DECam
images in a $\sim$1200 deg$^2$ region around the southern periphery of the Magellanic Clouds.


While scientific analysis with SMASH data have been published and are underway, we expect that there are many projects that others will think of and do first, as evidenced by the discovery of new star cluster candidates in the MCs by \citet{Piatti17} using SMASH data.  The aim of this paper is to enable more such projects by describing the SMASH survey, providing details on the pipelines that produced the photometric catalogs, and providing information on the  
first public data release.
The layout of this paper is as follows.  Sections \ref{sec:surveystrategy} and \ref{sec:observing} detail the survey and observing strategy.  An overview of the image
processing is given in Section \ref{sec:reduction}, the photometric reduction in Section \ref{sec:photometry}, while the calibration is discussed in 
Section \ref{sec:calibration}.  A description of the final catalogs and the achieved performance
is given in Section \ref{sec:performance}.  The details of the first SMASH data release and data access are
described in Section \ref{sec:dr1} and, finally, some of the first SMASH science results are discussed in Section \ref{sec:results}.




\section{Survey Strategy}
\label{sec:surveystrategy}

Figure \ref{fig_bigmap} shows the region of the sky that is relevant to the Magellanic Clouds and the Magellanic Stream\footnote{Also see Figure \ref{fig_dr1map} which has
more information on the SMASH fields including field names and calibration status.}, with the \hi distribution in the top panel
and the predicted stellar distribution of the \citet{Besla13} model in the bottom panel.  The DES footprint already covers one half of the LMC/SMC periphery as
well as much of the trailing Magellanic Stream.  The SMASH footprint was designed to cover the rest of the Magellanic periphery and the Leading
Arm\footnote{Note that even though the SMASH fields were designed to be complementary to the DES survey, the DES footprint changed over the last couple of
years producing some overlap and gaps between the two surveys.}, but
avoid the Milky Way mid-plane that could ``contaminate" the data.  A fully-filled survey would have been preferred, but to reach the sensitivity required to detect the
predicted low surface brightness features would have required hundreds of nights and would be beyond the possibility of an NOAO survey proposal.
We, therefore, decided to pursue a deep but partially-filled survey strategy as is shown in the top of Figure \ref{fig_bigmap} (red hexagons).  The SMASH fields map an
area of $\sim$480 deg$^2$ but are distributed over (and probe the stellar populations of) $\sim$2400 deg$^2$ with a filling factor of $\sim$20\%.

\begin{center}
\begin{deluxetable*}{lcccccccccccc}
\tabletypesize{\scriptsize}
\tablecaption{SMASH Fields Table}
\tablecolumns{13}
\tablewidth{500pt}
\tablehead{
  \colhead{Num} & \colhead{Name} & \colhead{RAJ2000} & \colhead{DEJ2000} & \colhead{RADEG} &
  \colhead{DEDEG} & \colhead{L$_{\rm MS}$\tablenotemark{a}} & \colhead{B$_{\rm MS}$\tablenotemark{a}} & 
  \colhead{$u$Flag} & \colhead{$g$Flag} & \colhead{$r$Flag} &  \colhead{$i$Flag} &  \colhead{$z$Flag} \\
}
\startdata
   1 &  0010-6947 &   00:10:19.8 &  $-$69:47:40.5 &     2.58282 &  $-$69.7946 &   $-$19.56937 &   $-$13.84173 & 4 & 4 & 4 & 4 & 4 \\
   2 &  0018-7705 &   00:18:57.9 &  $-$77:05:00.2 &     4.74128 &  $-$77.0834 &   $-$12.11547 &   $-$15.01799 & 4 & 4 & 4 & 4 & 4 \\
   3 &  0023-7358 &   00:23:19.0 &  $-$73:58:08.0 &     5.82950 &  $-$73.9689 &   $-$15.14808 &   $-$13.94466 & 1 & 1 & 1 & 1 & 1 \\
   4 &  0024-7223 &   00:24:56.6 &  $-$72:23:08.1 &     6.23604 &  $-$72.3856 &   $-$16.67737 &   $-$13.38564 & 1 & 1 & 1 & 1 & 1 \\
   5 &  0044-7137 &   00:44:06.7 &  $-$71:37:45.8 &    11.02820 &  $-$71.6294 &   $-$16.91433 &   $-$11.74021 & 1 & 1 & 1 & 1 & 1 \\
   6 &  0044-7313 &   00:44:32.5 &  $-$73:13:31.7 &    11.13560 &  $-$73.2255 &   $-$15.38135 &   $-$12.28828 & 1 & 1 & 1 & 1 & 1 \\
   7 &  0045-7448 &   00:45:03.2 &  $-$74:48:14.7 &    11.26370 &  $-$74.8041 &   $-$13.85882 &   $-$12.82186 & 1 & 1 & 1 & 1 & 1 \\
   8 &  0050-8228 &   00:50:34.5 &  $-$82:28:26.4 &    12.64380 &  $-$82.4740 &    $-$6.36406 &   $-$15.28088 & 1 & 1 & 1 & 1 & 1 \\
   9 &  0101-7043 &   01:01:27.4 &  $-$70:43:05.5 &    15.36420 &  $-$70.7182 &   $-$17.19874 &   $-$10.09455 & 1 & 1 & 1 & 1 & 1 \\
  10 &  0103-7218 &   01:03:36.3 &  $-$72:18:54.3 &    15.90130 &  $-$72.3151 &   $-$15.66017 &   $-$10.63165 & 2 & 2 & 2 & 2 & 2 \\
\enddata
\tablenotetext{0}{Full table is available in the electronic version of this paper and in the {\tt data/smash\_fields\_final.txt} file on the
SMASHRED repository (\url{https://github.com/dnidever/SMASHRED}).}
\tablenotetext{1}{Magellanic Stream coordinates defined in \citet{Nidever08}.}
\label{table_smashfields}
\end{deluxetable*}
\end{center}

The DECam imager is composed of 62 chips (59 and a half were functional throughout most of the SMASH observing; CCDNUM 2, 61 and one amplifier of 31 are were
not producing useful science data although CCDNUM 2 started working again as of December 29, 2016)
from Lawrence Berkeley National Lab (LBNL)
arranged in a hexagonal pattern covering a field of view of $\sim$3 deg$^2$ and a width of $\sim$2\degr.
The SMASH fields were chosen using an all-sky tiling scheme in which we laid down a uniform hex pattern of field centers with 1.7\dgr separation between field
centers with coordinates based on an Interrupted Mollweide projection.  This spherical projection has low distortion, such that a uniform sampling in its coordinate system
produces tiling with few areas of excessive overlap between fields.  We then transformed the coordinates of the hex-based tiling to spherical coordinates, and
rotated the coordinate system to place the seams and poles (southern pole of [$\alpha$,$\delta$]=[10\degr,$-$30\degr]) in areas outside of our survey area.
The resulting tiling of the sky was nearly uniform over our survey area with $\sim$15\% areal overlap between neighboring fields to allow for good cross-calibration
(although this was only used in our main body fields).

From this list of tiles we selected 154 fields by hand to uniformly sample the region of interest with a $\sim$20\%
filling factor as well as fully cover the inner regions of the LMC and SMC.  The full coverage tiling scheme and overlap was used so
that, given more observing opportunities, we could more easily completely cover regions with interesting stellar populations later on, which is what we did for the outer LMC disk.
Note that the final survey tiling scheme was created after the 2013, March 17--20 pre-survey run. Therefore, the positions of the 23 Leading Arm fields (Fields 153 -- 175) that were
observed on that run are not entirely consistent with those of the final tiling scheme, but the differences are not significant.
The final list of SMASH fields with coordinates in various systems is
available in electronic version of Table \ref{table_smashfields}.

\section{Observing Strategy and Observations}
\label{sec:observing}

The idea for SMASH was conceived during the NOAO ``Seeing the Big Picture: DECam Community Workshop" in Tucson, AZ on August 18--19, 2011.
We decided to submit a proposal for a Magellanic Clouds pilot project using Science Verification (SV) and Shared Risk (SR) time during the 2012B season.
The goal of the successful project (SV:2012B-3005 and SR:2012-0416) was to ascertain the necessary filters and depth to attain the needed sensitivity to Magellanic stellar
populations. Data were obtained in five fields at various distances from the Magellanic Clouds and included exposures in all five $ugriz$ bands and to a depth $\sim$1 mag
deeper than we thought was necessary for our science goals.  These data helped us evaluate various observing and survey strategies.  Ultimately, it was determined
that all five bands would give us the sensitivity needed to detect the predicted stellar populations in the Magellanic periphery and the Leading Arm.  In addition, the $u$-band
would allow us to determine photometric metallicities for the Magellanic main-sequence stars \citep[e.g.,][]{Ivezic08} which would be useful in determining the origin of any
newly-found stellar populations.


After the pilot project, there was no call for survey proposals, so we proceeded to submit a regular NOAO proposal to look for stellar populations in the area of the
Leading Arm (2013A-0411).  To maximize the coverage we did not take $u$-band exposures for this observing run; however, the $u$-band exposures were obtained
on later observing runs.

There was a call for survey proposals during the next semester, and we submitted a successful proposal for the SMASH survey of the Magellanic Cloud stellar populations
(2013B-0440).  We were originally awarded 30 DECam nights (with a 7/3 A/B semester split) and 14 0.9-m nights for calibration purposes over three years.
The standard SMASH observing sequence for a science field is three 60 s exposures (with large, half chip offsets) in each band and three deep exposures with exposure
times of 333 s ($u$), 267 s ($g$), 267 s ($r$), 333 s ($i$), and 333 s ($z$) with small $\sim$2\arcsec~dithers.  Each field takes about 110 min to observe including readout time and slewing.
Each night exposures of four to five standard star fields (focusing on the \citealt{York00} SDSS equatorial region where data for all chips could be obtained simultaneously)
were obtained with exposure times of 1 s in all $ugriz$ bands as well as 10 s in $griz$ and 60 s in $u$, but to save time this was later changed (half-way through the survey)
to single exposures of 15 s in $griz$ and 60 s in $u$.

\begin{center}
\begin{deluxetable*}{llll}
\tablecaption{SMASH DECam and 0.9-m Observing Runs}
\tablecolumns{4}
\tablewidth{450pt}
\tablehead{
\colhead{Date (nights)} & \colhead{Telescope} & \colhead{Source} & \colhead{Comments}
}
\startdata
\sidehead{Pre-Survey}
\hline4-m
Dec 11+12, 2012 (2)     & 4-m    & Shared Risk  &  5 pilot fields \\
Mar 17--20, 2013 (4)      & 4-m  & 2013A-0411  &  23 fields ($griz$)  \\
Aug 8+9, 2013 (2 part)      &  4-m    & Time from Saha Bulge project & clear, 3 fields\hspace{4.5cm}  \\
\hline
\sidehead{Survey Year 1}
\hline
Jan 5--7, 2014 (3)     &  4-m  & NOAO survey  & 0.5 night lost, 10 fields \\
Jan 12+13, 2014 (2)    &  0.9-m  & Makeup for Oct 21+22  & 2 nights photometric, 4 fields calibrated \\
Jan 19+20, 2014 (2 half)    &  4-m    & DD time  & clear, $riz$ for 6 fields\\
Jan 21--28, 2014 (8 half) &  4-m    & Chilean time  &  1 half night lost, 4 fields, 9 partials \\
Jan 29+30, 2014 (2 half)    &  4-m    & DD time  &  clear, $ug$ for 8 pre-survey fields\\ 
Feb 13, 2014  (1 part)  &  4-m    & Engineering  & clear, $riz$ for 6 fields \\
Feb 14--23, 2014 (10)  &  0.9-m  & NOAO survey &  9 nights photometric, 30 fields calibrated\\
May 27--June 2, 2014 (7)  & 4-m & NOAO survey & lost 1 night, 21 fields observed, $ug$ for \\
& & & 13 pre-survey fields, 3 extra fields \\
\hline
\sidehead{Survey Year 2}
\hline
Sep 25--Oct 1, 2014 (7)  & 0.9-m  & NOAO survey &  1 night photometric, 11 fields calibrated \\
Oct 11--12, 2014 (2)  &  4-m  &  Engineering  &  some globular cluster calibration data  \\ 
Nov 21--23, 2014 (3)  &  4-m  &   NOAO survey &  12 LMC/SMC main-body fields \\
Dec 17--18, 2014 (2)  &  4-m  &   NOAO survey &  8 LMC/SMC main-body fields \\
Mar 13--18, 2015 (5) &  4-m  &   NOAO survey &  mostly clear, 21 finished, 4 partials \\
Mar 30--31, 2015 (2)   &  4-m    & DD time  & deep \& high-cadence data of Hydra II \\ 
Apr 26--Mar 2, 2015 (7) & 0.9-m &  NOAO survey & 4.5 nights photometric, 48 fields calibrated \\
\hline
\sidehead{Survey Year 3}
\hline
Nov 9, 2015 (1)   &  4-m &  NOAO survey  &  clear, 4 fields  \\
Nov 23, 2015 (1)        &  4-m       & DD time &  bad weather, long $riz$ for 2 fields \\
Nov 27--29, 2015 (3)    &  0.9-m     &  Chilean time & 9 fields calibrated \\
Dec 5+6, 2015 (2)  & 4-m &  NOAO survey & 8 fields, 7 are LMC/SMC main-body \\
Jan 1--6, 2016 (6)  &  4-m  &   NOAO survey   &  4 nights lost, 3 finished, 2 partials  \\
Feb 13--18, 2016 (6) & 4-m  &  NOAO survey  &  40 shallow LMC fields, 18 long fields \\
\hline
\sidehead{Survey Year 4 -- Extension}
\hline
Oct 29--31, 2016 (3)    &  4-m       & NOAO survey &  0.5 night lost, 8 fields
\enddata
\label{table_observations}
\tablenotetext{1}{Full up-to-date table is available in  the {\tt obslog/smash\_observingruns.txt} file on the
SMASHRED repository (\url{https://github.com/dnidever/SMASHRED).}}
\end{deluxetable*}
\end{center}

Due to bad weather, poor seeing (we have seeing constraints of $\lesssim$1\arcsec~for the central LMC/SMC main-body fields because of crowding), and the short B semester nights, the survey
fell behind in the MC main-body regions.  Therefore, after the first year we requested our 10 nights per year be split evenly between the A and B semesters (instead of 7/3 as before),
and after the second year, we requested an additional three nights per semester in 2015B and 2016A.  After our last year, we requested a three night extension in
2016B to fill a ``hole" in our coverage of the SMC periphery (near the south celestial pole) of 11 fields.  Additional DECam nights were obtained through the Chilean
TAC (PI: Mu\~noz; 2014 Jan.\ 21--28).  Finally, after the discovery of the Hydra II Milky Way satellite in the SMASH data \citep{Martin15}, we submitted a
Director's Discretionary Time proposal to obtain time-series data on Hydra II to study variable stars (2015 March 30--31).

On our very successful 2016 Feb 13--18 run, we finished all of the fields around the Magellanic Clouds that were observable and, therefore, we decided to
observe some ``extra'' shallow fields around the LMC that would help reveal structures in the LMC disk (similar to those seen by \citealt{Mackey16} and
\citealt{Besla16}) and allow us to use the field overlaps to create a more homogeneously calibrated dataset around the LMC using an ``\"ubercal" technique \citep[e.g.,][]{P08}.
Two 60 s exposures in $griz$ with a one half-chip dither between the pairs were obtained for these 40 fields (with field numbers between 184 and 243).
The last SMASH observing run in the Leading Arm region (2016 May 8--12) was completely lost due to bad weather, and, therefore, the fields in that
region were not completed.

SMASH was allotted 57 nights of DECam observations (on 63 separate nights with 12 of these being half nights) and 47 nights (or 75\%) of useful data were obtained.
All allocated observing for the SMASH survey has now concluded.
The median seeing in $ugriz$ is (1.22\arcsec, 1.13\arcsec, 1.01\arcsec, 0.95\arcsec, 0.90\arcsec), respectively, with a standard deviation of $\sim$0.25 and 
the median airmass of all observations is 1.35.  Useful exposures were obtained for 197 fields (158 deep fields) but 27 fields
from our original survey plan remain unobserved due to poor weather conditions (mainly in the Leading Arm region).
Table \ref{table_observations} shows all SMASH survey observing time including all time and data from non-NOAO sources (which are all combined as part of the SMASH dataset).
More information about which nights were photometric are in the {\tt smash\_observing\_conditions.txt} file (see Section \ref{calib:transphot} below).

\subsection{0.9-m Observations}
\label{subsec:0.9mobservations}

The CTIO 0.9-m telescope and Tek2K CCD camera were used to collect observations of SDSS standards and SMASH fields in order to provide an independent calibration of a portion of the DECam data, particularly for fields observed under non-photometric conditions with the 4-m telescope.  The bulk of these observations were taken using CTIO's SDSS $ugriz$ filter set, while for three nights we used the borrowed DES PreCam $griz$ filters\footnote{The DES PreCam filters are a 4$\times$4 inch filter set made to the same specification as the full-sized DECam filters.} for the sake of comparison.  

The typical nightly observing plan in photometric weather was to observe several standard star fields from Smith et al. (2002) and from SDSS Stripe 82 and Stripe 10 \citep[DR12;][]{DR12} at the beginning and end of each night and every $\sim$2 hours in between, and observe SMASH fields during the rest of the time.  The Tek2K camera has a 13.5\arcmin$\times$13.5\arcmin~field of view, and so covered only the central portion of the SMASH fields.  We did not offset the 0.9-m to cover the full DECam field of view, and so obtained calibration information only for the central DECam chips.  Typical exposure times for the standard fields were 300 s ($u$), 20 s ($g$), 5 s ($r$), 10 s ($i$), and 15 s ($z$), while for the SMASH fields we took sets of five undithered exposures with individual exposure times of 600 s ($u$), 60 s ($g$), 60 s ($r$), 120 s ($i$), and 360 s ($z$).  During non-photometric 0.9-m nights, we only took images of SMASH fields, and used short exposures of these fields taken on photometric nights to bootstrap the calibration of the non-photometric exposures.  Table \ref{table_observations} also summarizes the 0.9-m observing runs.

Calibration data taken at the telescope consist of daily dome flats in $griz$, twilight sky flats in $ugriz$, exposures for the creation of a shutter shading map, and exposures for the creation of a bad pixel mask.  The shutter shading calibration data consist of $r$-band dome flats observed while repeatedly opening the shutter for one second and closing it during the exposure, intermingled with normal dome flats taken with the same total exposure time as the shutter frames.  The bad pixel mask data consisted of
100 0.1-s $r$-band dome flat exposures and a set of 6 $r$-band dome flats taken with levels equaling 75\% of saturation.

\section{Image Processing}
\label{sec:reduction}

This section describes the image processing from raw to flattened and ``detrended" final images for both the DECam and 0.9-m data.

\subsection{Image Processing of the DECam Data}
\label{subsec:decamred}

The SMASH data reduction of the DECam data makes use of three separate software packages: (1) the NOAO Community Pipeline
\citep[CP;][]{Valdes2014}\footnote{\url{http://www.noao.edu/noao/staff/fvaldes/CPDocPrelim/PL201\_3.html} also see the NOAO Data Handbook
\url{https://www.noao.edu/meetings/decam/media/DECam\_Data\_Handbook.pdf}} for instrument signature removal (Section \ref{subsubsec:cpred}),
(2) PHOTRED\footnote{\url{https://github.com/dnidever/PHOTRED}} for PSF photometry (Section \ref{photred}), and
(3) SMASHRED, custom software written for PHOTRED pre- and post-processing of the SMASH data (Section \ref{subsubsec:smashredpreproc} and \ref{subsec:calibsoftware})).

\subsubsection{Community Pipeline Reductions}
\label{subsubsec:cpred}
The CP was developed by the Dark Energy Survey Data Management (DESDM) team but with subsequent important modifications by F. Valdes at NOAO
to produce reduced images for the community.
The CP performs the following operations on the data:
\begin{itemize}
\item Bias correction.
\item Crosstalk correction.
\item Saturation masking.
\item Bad pixel masking.
\item Linearity correction at both low and high count levels.
\item Flat field calibration.
\item Fringe pattern subtraction, for $z$ and $Y$ bands.
\item Bleed trail and edge bleed masking and interpolation.
\item Astrometric calibration of the image WCS with 2MASS as the astrometric reference catalog.
\item Single exposure cosmic ray masking, by finding pixels that are significantly brighter than their neighbors.
\item Photometric calibration using USNO-B1.
\item Sky pattern removal.  The ``pupil ghost" and spatially varying background are subtracted.
\item Illumination correction using a ``dark sky illumination" image.
\item Remapping to a tangent plane projection with constant pixel size.
\item Transient masking with multiple exposures.
\item Single-band coadding of remapped exposures with significant overlap.
\end{itemize}

The CP is run by NOAO staff on all the community DECam data.  It generally takes a week after the completion of an observing run for the data to be processed
and become available on the NOAO Science Archive.\footnote{\url{https://www.portal-nvo.noao.edu}} The CP produces instrumentally calibrated images (``InstCal"), remapped
versions of InstCal (``Resampled"), and single-band coadded images (``Stacked").  For SMASH we use the InstCal images which come in three multi-extension (one per chip) and
fpack\footnote{\url{https://heasarc.gsfc.nasa.gov/fitsio/fpack/}} compressed FITS files per exposure: flux (``image"), weight/variance (``wtmap"), and quality mask (``qmask").

\subsection{Image Processing of the 0.9-m Data}

We used the NOAO/IRAF QUADRED package, custom IDL programs, and other software to process the images from the 0.9-m observations.  The basic steps of this processing are:
\begin{enumerate} 
\item Electronic crosstalk correction, using custom software to measure and correct for the electronic ghosting present in the images when read through multiple amplifiers.
\item Correction for electronic bias using the CCD's overscan region and bias frames.
\item Trimming of the images to the illuminated area.
\item Derivation of the exposure time-dependent illumination map caused by the opening and closing of the camera's iris shutter using dome flat observations designed for the purpose, and application of this shutter shading correction to the observed images.
\item Derivation of flat field frames from twilight sky images and application to the object frames.
\item Derivation of a bad pixel mask from dome flat observations designed for the purpose, with bad pixel correction applied to the object frames.
\item Derivation of dark sky flats ($ugri$) and fringe frames ($z$) by stacking and filtering the deep sky observations taken throughout each observing run, followed by division by the dark sky flats ($ugri$) and subtraction of fringe features ($z$) for all object frames.
\item Use of the code library from http://astrometry.net (Lang et al.\ 2010) to populate the object image headers with World Coordinate System (WCS) solutions.
\end{enumerate}

\section{Photometric Reduction}
\label{sec:photometry}

This section describes how photometric catalogs were created from the reduced images of the DECam and 0.9-m data.

\subsection{DECam Pre-Processing with SMASHRED}
\label{subsubsec:smashredpreproc}

The CP-reduced images are not in a format that is readable by DAOPHOT.  Therefore, we run a SMASH pre-processing script 
({\tt SMASHRED\_PREP.PRO}) on the CP images for each night before PHOTRED is run.  This script performs the following steps:
\begin{enumerate}
\item Rename files in the old (``tu") naming convention to the new (``c4d") convention (the official NOAO archive file naming
convention\footnote{\url{http://ast.noao.edu/sites/default/files/File\_Naming\_Conventions\_v12.pdf}} was changed in early 2015).
\item Move standard star exposures to the ``standards/" directory since they are processed separately from the science data (see Section \ref{subsubsec:stdred}).
\item Uncompress the FITS files, set ``bad" pixels to 65,000, and write new FITS files for each chip image.
\item Sort the exposures into PHOTRED ``fields" based on the pointing and exposure times (short and long exposures are processed separately).  Rename the files
using the PHOTRED file naming convention (\verb|FIELD#-EXPNUM#_CHIP#.fits|, e.g., \verb|F5-00507880_17.fits|).
\item Download astrometric reference catalogs for each field and write separate reference catalog files for each chip FITS file.
\item Move files for each field into a separate directory (e.g., {\tt F8/}, {\tt F9/}, {\tt F10/}, {\tt F11/}).
\end{enumerate}

The CP mask files provide information on bad pixels, saturation, bleed trail, cosmic rays, and multi-exposure transients.  There were some
problems with the multi-epoch transient masking so we ignored that information in the mask.  Any pixels that were affected by the other
issues were set to a high value (65,000) so that PHOTRED/DAOPHOT would see these pixels as ``bad" and ignore them.

\subsection{DECam Nightly PHOTRED Reduction}
\label{photred}

Accurate, point-spread-function (PSF) fitting photometry was obtained using the automated PHOTRED pipeline first described in \citet{Nidever11}.  PHOTRED performs
WCS fitting, single-chip PSF photometry as well as multi-exposure forced-PSF photometry using the DAOPHOT suite of programs \citep{Stetson87,Stetson94}.
PHOTRED was run separately on each night.  The short and long exposures of a field were run through PHOTRED separately (the former with a ``sh" suffix
added to their name) and multi-band image stacking and forced photometry were only performed on the long exposures.  This was mainly because of issues with
bright, saturated stars when stacking short and long exposures and the fact that the short exposures did not add much to the overall depth of the longer exposures.
Note also that deep exposures of a field taken on different nights were processed separately and only combined during the calibration stage (see Section \ref{sec:calibration}).

PHOTRED is based on methods and scripts developed by graduate students and postdocs in S. Majewski's ``halo" group at the University of Virginia (UVa) in the late
1990s and early 2000s (in particular J. Ostheimer, M. Siegel, C. Palma, T. Sohn \& R. Beaton).  PHOTRED fully automates these scripts
(and some manual procedures) into a robust and easy-to-use pipeline.  Most of the PHOTRED software was written by D.L.\ Nidever in 2008 while he was a graduate
student at UVa and has been continually updated and improved since then.  PHOTRED consists of IDL\footnote{The Interactive Data Language is a product of
Exelis Visual Information Solutions, Inc., a subsidiary of Harris Corporation.} driver programs wrapped around the DAOPHOT Fortran routines,
but also includes some IRAF, Fortran and Unix shell scripts.

PHOTRED currently has 13 ``stages".  Text-based lists are used for keeping track of inputs, outputs and failures and shuffling files from one stage to the next.
This overall design was taken partly from the SuperMACHO ``photpipe" pipeline \citep{Rest05,Miknaitis07}.  The global parameters and optional settings (see the github
repository for the full list) as well as the stages to be run are specified in the {\tt photred.setup} setup file.   The majority of the stages work on a chip-by-chip level
(SPLIT--CALIB) and these chip-level catalogs are combined for the few final stages (COMBINE--SAVE).
Although not all are used for SMASH, for completeness and future reference all of the stages are described in detail below.

\subsubsection{RENAME}
The headers are checked for all the required keywords (gain, read noise, time stamp, filter, exposure time, $\alpha$/$\delta$, airmass).  The exposures are grouped
into ``fields" based on values in the ``object" keyword in the header and renamed with the PHOTRED naming convention (\verb|FIELD#-EXPNUM#_CHIP#.fits|).  The
PHOTRED short field names and full field names are saved in the {\tt fields} file. This stage was
skipped for SMASH since it is already performed by the {\tt SMASHRED\_PREP.PRO} pre-processing script.

\subsubsection{SPLIT}
If the FITS files are multi-extension files then these are split into separate FITS files per chip.  This stage was also skipped for SMASH.

\begin{figure}[t]
\includegraphics[width=0.97\hsize,angle=0]{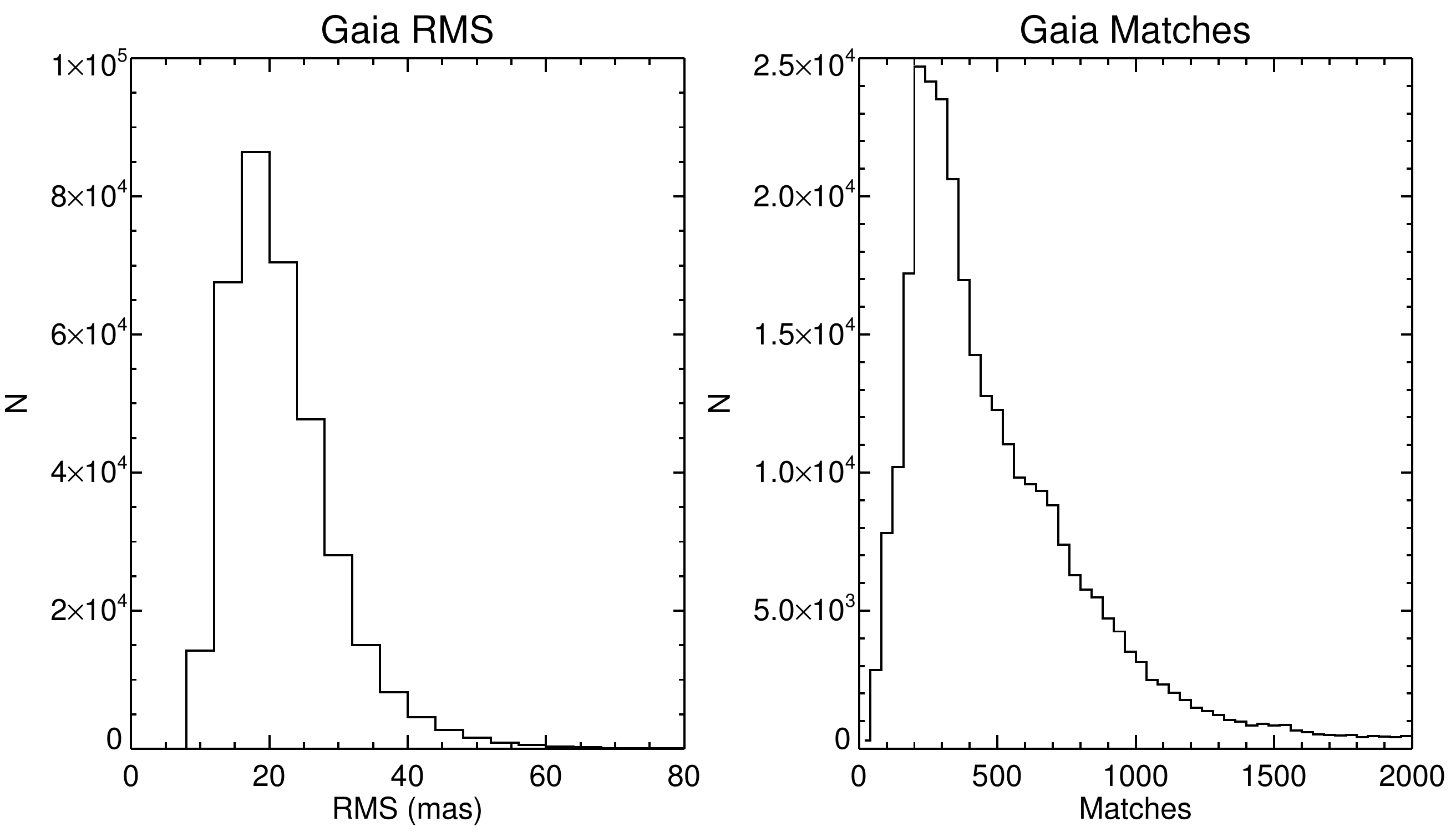}
\caption{\small
(Left) The distribution of rms values between the Gaia astrometric reference catalog and the
SMASH DECam data for the final WCS (for the $\sim$350,000 separate SMASH chip files).
(Right) The distribution of matches per chip between the SMASH and Gaia data with a median of $\sim$500 matches per chip.
}
\label{fig_gaiarms}
\end{figure}

\subsubsection{WCS}
\label{subsubsec:wcs}
The world coordinate system (WCS) for an image is created (or refined if it already exists in the header) by using an astrometric reference catalog and some
information about the imager (pixel scale and orientation) and pointing (rough $\alpha$/$\delta$ of the center of the image, normally from the Telescope Control System).
The software ({\tt WCSFIT.PRO}) performs its own simple source detection, sky estimation and aperture photometry of the image using routines from the
IDL Astronomy User's Library\footnote{\url{http://idlastro.gsfc.nasa.gov}}.  If a WCS does not already exist, then the reference catalog $\alpha$/$\delta$ values
are transformed roughly to the $x/y$ cartesian coordinates of the image by using the exposure and image information provided.  The reference sources are then
cross-matched with the image sources by cross-correlating down-sampled ``detection" map images of the two groups of sources.  The peak in the cross-correlation
image is used to obtain an initial measurement of the $x/y$ offsets between the lists and the significance of the match.  If a significant match is found then
nearest-neighbor matching is performed with a large matching radius and the measured offsets.  The matches are used to fit a four parameter transformation
matrix (essentially translation, rotation and scale) and a second round of improved nearest-neighbor matching.  The final matches are used to perform fitting
of the four \verb|CD#_#| and two \verb|CRVAL#| parameters of the WCS.  The software does not create or modify existing higher-order distortion terms
but uses if them if they already exist in the header (e.g., for DECam the existing \verb|PV#_##| TPV distortion terms are used).

By default, the SMASHRED pre-processing used USNO-B1\footnote{\url{http://tdc-www.harvard.edu/catalogs/ub1.html}} \citep{Monet03} as the astrometric
reference catalog, and sometimes 2MASS or UCAC4 \citep{Zacharias03}.  The rms of the residuals of the astrometric solutions with these catalogs gave values of $\sim$270 mas.
After the first Gaia data release \citep{GaiaDR1}, the WCS-fitting
software was rerun with Gaia as the astrometric reference and the resulting FITS header (with the improved WCS) saved in a separate text file ({\tt gaiawcs.head})
for each image.  The astrometric solutions were dramatically improved, with a resultant rms of only $\sim$20 mas (see Figure \ref{fig_gaiarms}).

\begin{deluxetable}{ll}
\tablecaption{Default PHOTRED Options for DAOPHOT}
\tablecolumns{2}
\tablehead{
\colhead{Option} & \colhead{Comment}
}
\startdata
TH = 3.5$\sigma$  & Detection threshold \\
VA = 2  &    Quadratic spatial PSF variations \\
FI = 1$\times$FWHM  &  PSF fitting radius \\
AN = $-$6   &  Use lowest $\chi^2$ analytical PSF model
\enddata
\label{table_options}
\end{deluxetable}

\begin{figure*}[t]
\begin{center}
\includegraphics[width=0.97\hsize,angle=0]{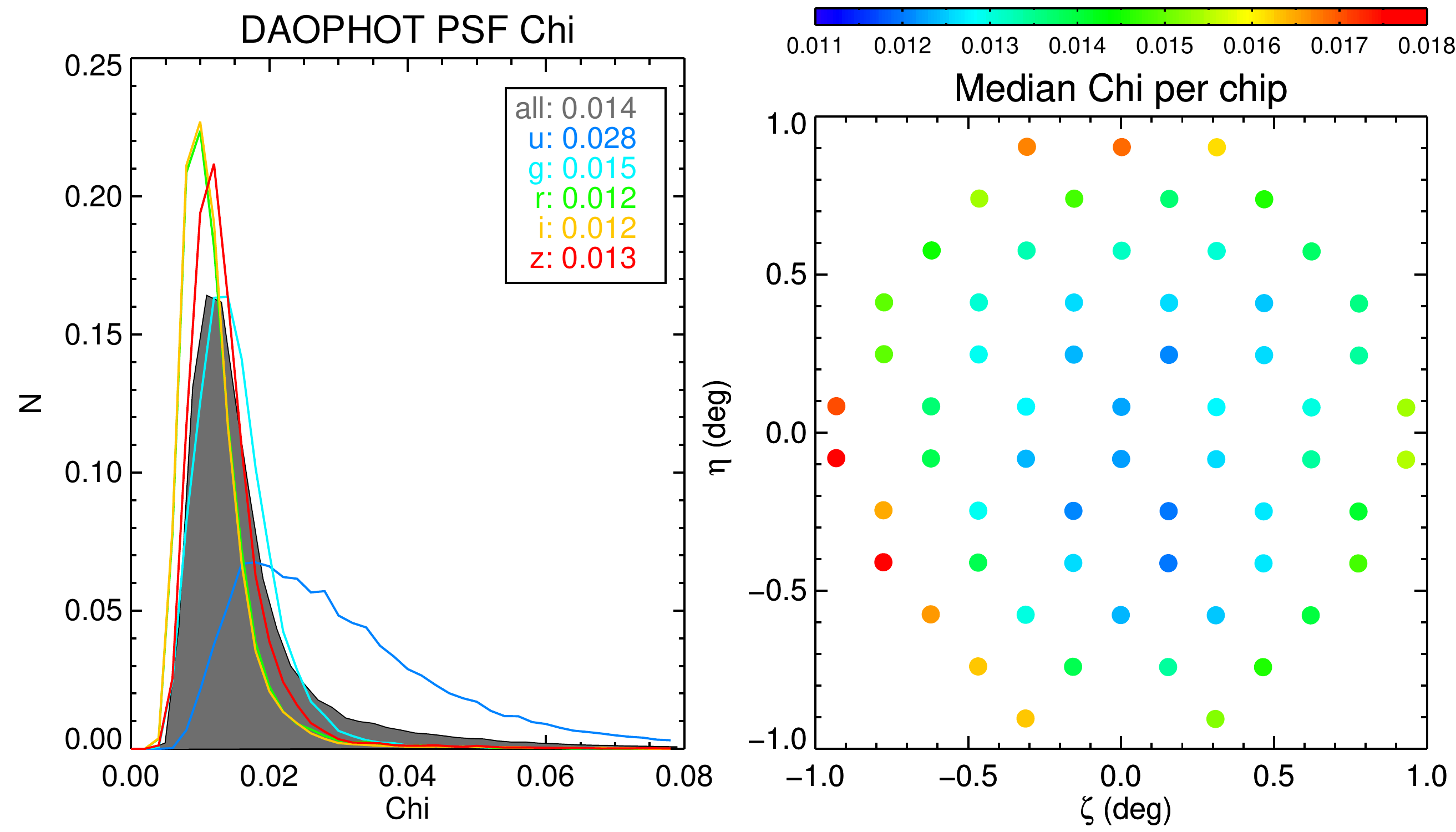}
\end{center}
\caption{\small
(Left) The distribution of DAOPHOT PSF ``chi" values (relative root-mean-square of the analytic PSF residuals) for the $\sim$350,000 separate SMASH chip files
broken down by band.  The median values per band are given in the legend. (Right) The median chi value per chip (over all bands) as they appear in the focal plane.}
\label{fig_psfchi}
\end{figure*}

\subsubsection{DAOPHOT}
\label{subsubsubsec:daophot}

This stage detects sources in the single exposure images, constructs the PSF, and uses it to measure PSF photometry with ALLSTAR.
There are several steps:

\begin{enumerate}
\item {\bf PSF FWHM Estimate:} Since DAOPHOT requires an estimate of the PSF FWHM (full width at half maximum, or seeing), a custom IDL routine ({\tt IMFWHM.PRO)}
with independent algorithms is used to make this estimate.  The routine detects peaks in the image 8$\sigma$ above the background (although this is lowered if
none are detected) and keeps only those peaks having the maximum value within 10 pixels of the peak position and have two or more neighboring
pixels that are brighter than 50\% of the peak's flux (to help weed out cosmic rays).  The routine then finds
the contour at half-maximum flux in a 21$\times$21 sky-subtracted sub-image centered on the peak and uses it to measure an estimate of the FWHM
(2$\times$ the mean of the radius of the contour) and the ellipticity of the contour.  In addition, the total flux in the subimage and a ``round" factor
(similar to the DAOPHOT version) using marginal sums are computed.  These metrics are then used to produce a cleaner list of sources (FWHM$>$0, round$<$1,
ellipticity$<$1 and flux$<$0) on which two-dimensional Gaussian fitting is performed and more reliable metrics are computed.  The final list of sources
is selected by cuts on the new metrics and the distributions of semi-major axis, semi-minor axis and $\chi^2$ (selected sources must lie within the dominant
clustering of these parameters for all sources).  The final FWHM and ellipticity are then computed from these sources using robust averages with outlier
rejection.  This FWHM value is then used in the next step to set the DAOPHOT input options.

\item {\bf DAOPHOT option files:} Both DAOPHOT and ALLSTAR require option files ({\tt .opt} and {\tt .als.opt} respectively).
Some of the most important default settings are shown in Table \ref{table_options}.

For some very crowded fields (e.g., Field35 and Field46) the default settings produced suboptimal results and, therefore, linear PSF spatial variations ({\tt VA}=1)
and a smaller fitting radius ({\tt FI}=0.75$\times$FWHM) were used.  The affected nights are 20141123, 20141217, 20151205 and 20151206.

\item {\bf Common sources list:}  
Early on in the development of PHOTRED there were issues with constructing good PSFs for the deep (280 s),
intermediate-band DDO51 observations for the MAPS survey \citep[][which was the main motivation
for writing PHOTRED]{Nidever11,Nidever13}.  This was because there were a lot of point-like cosmic rays that overwhelmed
the small number of real sources and made it difficult to create a good PSF source list just by culling via morphological
parameters.  To deal with this problem, PSF sources were required to be detected in multiple exposures (across all filters) to make sure they were real objects.
In this step, a ``common sources list" is constructed for each FITS image and later used as the starting point to select PSF stars.
In the DECam data, the original issue is not as much of a problem because of the broad-band filters and because the
cosmic rays in the LBNL detectors tend to be more linear and less point-like in shape.  However, we have continued to use the common
sources option in the DAOPHOT stage for SMASH.

\item {\bf Detection:} Sources are detected in the images with {\tt FIND}, and aperture photometry is determined with {\tt PHOTOMETRY}
with an exponential progression of apertures from 3 to 40 pixels and sky radius parameters of 45 (inner) and 50 (outer) pixels.

\item {\bf Construct PSF:} The PSF is constructed with an iterative procedure for culling out ``suspect" sources.  The initial list of 100 PSF
sources is selected using {\tt PICK} from the common source list (or the aperture photometry file if the common source option was not used)
and a morphology cut is applied (0.2 $\leq$ sharp $\leq$ 1.0; using the sharp produced by {\tt FIND}) to remove extended objects.  The list is then
iteratively cleaned of suspect sources.  At each iteration a new PSF is constructed with {\tt PSF} using the new list and DAOPHOT prints out the
``chi" value (root-mean-square residual of the stellar profile from the best-fitting analytic model) for each star and flags any outliers
({\tt ?} and {\tt *} for 2 and 3 times the average scatter, respectively).  The flagged outliers and any sources with chi $>$ 0.5 are removed from
the list and the procedure is started over again until no more sources are rejected.  In most cases only one iteration is required.

After the list has converged, sources neighboring the PSF sources (using {\tt GROUP}) are removed from the image (using {\tt SUBSTAR}).
A new PSF is constructed from this ``neighbors subtracted" image and a similar iterative loop is used to remove PSF outlier sources.  As before,
most cases only require one iteration.  The median number of PSF stars used per chip image is $\sim$80.

The default PHOTRED setting is to allow DAOPHOT to pick the analytic function that produces the lowest chi value.  The most commonly
used function (77\% of the PSFs) is the four-parameter ``Penny", which is the sum of a Gaussian and Lorentz function.  The second most
common function (20\% of the PSFs) is a Moffat function with a power law exponent of $\beta$=3.5.

\item {\bf Run ALLSTAR:}
ALLSTAR is run to perform simultaneous PSF fitting on all the detected sources in the image using the constructed PSF.   The default
PHOTRED setting is to allow ALLSTAR to recentroid each source.  ALLSTAR is also run on the ``neighbors subtracted" images to obtain
PSF photometry for the PSF stars that are later used to calculate an aperture correction.  ALLSTAR outputs $x/y$ centroids, magnitudes with
errors, sky values, as well as chi and sharp (which describes how much broader the profile of the object appears compared to the profile of the PSF)
morphology parameters ({\tt .als} file).
\end{enumerate}

One of the failure modes for a file in this stage is not to have enough PSF stars after the cleaning to constrain the solution.  In these cases
the PSF spatial variation value ({\tt VA}) is lowered in the option file by hand and DAOPHOT rerun.  This solves the failures in the
large majority of cases. For the small number of files where this also fails, we select PSF sources by visual inspection.
The software will be modified to avoid these problems in the future by starting with a simple, constant analytic PSF and slowly add more complexity if it is needed.

\begin{figure*}[t]
\includegraphics[width=1.0\hsize,angle=0]{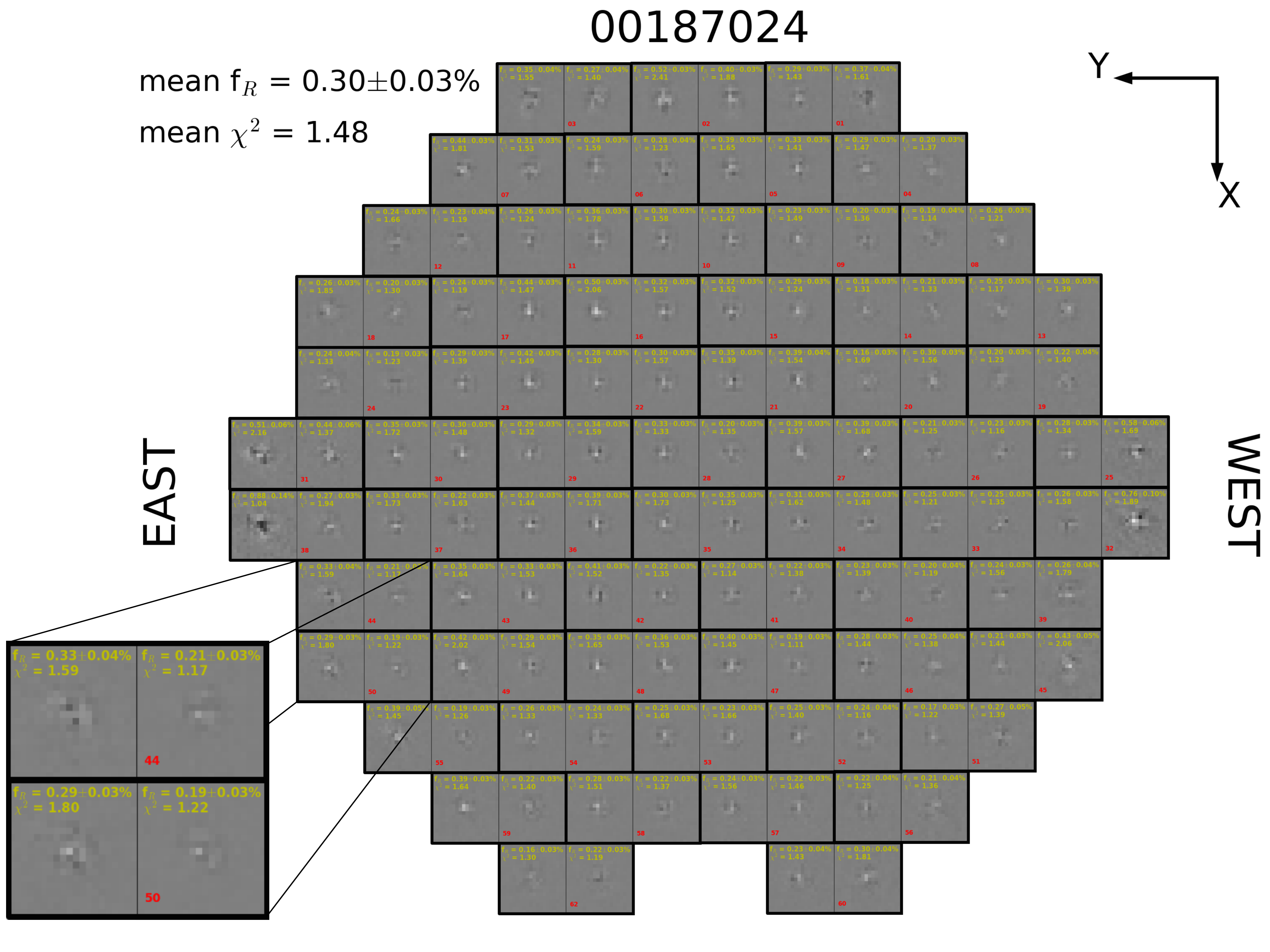}
\caption{PSF quality assurance figure for the full focal plane of a single exposure (EXPNUM=00187024) and a blow-up of two chips in the lower-left. Each square image shows
the relative residuals in the PSF-subtracted image (relative to total flux in the PSF model) medianed across $\sim$30 high $S/N$ stars (not necessarily PSF stars)
per half-chip.  Each horizontal rectangle
represents one chip of an exposure, and the two squares in the rectangle show the relative residuals for one half of the chip.  The relative absolute residuals and the uncertainties
(using propagation of errors from the noise in each image) as well as the $\chi^2$ for each half-chip are shown in the top of each square in yellow.
The range of the greyscale is $\pm$0.2\% and the chip number (CCDNUM) is shown in red.  Average values for the entire exposure are in the upper-left.}
\label{fig_qapsf}
\end{figure*}

Figure \ref{fig_psfchi} shows the histogram of DAOPHOT analytic PSF chi values broken down by band.
The $griz$ chi values are tightly peaked around $\sim$0.012 (or 1.2\%) while the $u$-band values are a factor of 2$\times$ larger.
The higher $u$-band values are because the $S/N$ of the PSF stars are on average lower, which gives rise to larger scatter in the residuals from photon noise.
The right-hand panel of Figure \ref{fig_psfchi} shows the median chi value per chip (across all exposures) as they appear on the sky, indicating that the analytic
first approximations are slightly poorer for the chips on the periphery of the focal plane, possibly due to optical distortions or non-conformity of the detectors
to the focal plane.  Note that any systematic differences
between the ``true" PSF and the analytic first approximation go into DAOPHOT's PSF empirical look-up table of corrections, so the final rms of the PSF residuals
will actually be smaller than the chi values.  Figure \ref{fig_qapsf} shows diagnostic thumbnails of median-combined relative flux residuals from PSF-subtracted
images of many bright stars.  The PSF relative flux error is of the order of $\sim$0.3\% with very little systematic structure left in the medianed residual image
indicating that the PSFs are of high quality.

\subsubsection{MATCH}
The sources in the ALLSTAR photometry catalogs from the DAOPHOT stage are cross-matched and combined with the files for each chip being handled
separately (i.e., all the chip 1 files are cross-matched together and the chip 2 files are cross-matched together, etc.).  Astrometric transformations between the frames
(using the $x/y$ cartesian coordinates) and a reference frame (which is chosen based on the longest exposure time frame in the {\tt filtref} band) are computed
(in similar manner to what DAOMATCH achieves).  The WCSs in the FITS headers are used to calculate an initial estimate of the transformations.
If this fails, then a more general matching routine is run that uses a cross-correlation technique of a down-sampled ``detection" map image
between the two source lists (as described above in Section \ref{subsubsec:wcs}).  Once all the transformations are in hand, DAOMASTER is used to
iteratively improve the transformations (written to {\tt .mch} file) and cross-match, and, finally, combine all of the photometry (per chip) into one merged file (the {\tt .raw} file).

\subsubsection{ALLFRAME}
PSF photometry can be improved by using a stacked image for source detection and then holding the position of the sources fixed while
extracting PSF photometry from each image.  The improvement of this ``forced" photometry over regular PSF photometry performed separately from
each image comes from the reduced number of free parameters (i.e., the positions).  PHOTRED makes use of the DAOPHOT ALLFRAME \citep{Stetson94}
program to perform the forced photometry.  The ALLFRAME stage processes each group of chips separately (e.g., chip 01 files separately from chip 02 files, etc.).

This stage performs several separate tasks:
\begin{enumerate}

\item {\bf Construct multi-band coadd:} A weighted average stack is created of all the images.  First, the relative flux scaling, sky level, and weights are computed for all the images.
The weights are essentially the $S/N$ and are based on sources detected in all of the images (if no sources are detected in all the images, then a bootstrap approach
is used to tie the images to one another).  Second, images are transformed to a common reference frame.  The original code only applied $x/y$ translations
to the images.  However, this was insufficient for larger dithers where the higher-order optical distortions become important and the software was rewritten to fully resample
the images onto the final reference frame. The type of transformation used can be found in the {\tt ALFTILETYPE} column (``ORIG" or ``WCS") of the final {\tt chips} catalog/table.
Finally, the images are average combined using the IRAF routine IMCOMBINE with bad pixel masking and outlier rejection (sigma clipping).   The detector gain recorded in the
image header is maintained because the images are scaled to the reference exposure.  However, new read noise and sky values\footnote{Sky is needed in the combined images because
DAOPHOT uses it as part of its internal noise model.} for the combined image ({\tt \_comb.fits}) are computed using the weights, scalings and sky values.  It is
challenging to preserve the fidelity of bright stars when combining deep and shallow exposures.  This is one reason why it was decided to process short and long
SMASH exposures separately in PHOTRED.

\item {\bf PSF construction:}  The PSF of the combined image is constructed using the same routine as in the DAOPHOT stage.

\item {\bf Iterative source detection:} Source detection is performed iteratively in two steps. (1) Detect new sources in the working image (PSF subtracted after the first iteration)
with Source Extractor \citep[SExtractor, which works well for detecting faint sources;][]{Bertin96}
and incorporate these new detections into the master source list.  (2) ALLSTAR is run on the original image (with the previously-found PSF) using the
current master source list and all sources that converge are subtracted.  This two-step process is repeated for the desired number of iterations ({\tt finditer} option in {\tt photred.setup}).
The ALLSTAR output from the last iteration is used as the final master source list
({\tt \_comb\_allf.als}).  The default detection settings for SExtractor are: use a convolution filter, $>$1$\sigma$ detection threshold, and a minimum area of 2 pixels per source.
For SMASH only two iterations were used since we found that further iterations produced mainly spurious new detections.

\item {\bf Run ALLFRAME:} ALLFRAME is run on all the images using their respective PSFs and the master source list constructed in the previous step.
ALLFRAME uses the coordinate transformations between images from the {\tt .mch} file (in the MATCH stage), but computes its own small, high-order
geometric adjustments (we use the option of 20 terms or cubic in $x$ and $y$) to these during the fitting process (it slowly adds in the higher orders to keep the solutions
constrained).  We allow a maximum of 50 iterations in ALLFRAME after which it outputs catalogs ({\tt .alf}) with $x/y$ coordinates (in that image's reference frame),
photometry with errors and chi and sharp morphology parameters.
\end{enumerate}

After ALLFRAME has finished, the results for the individual images are combined and the SExtractor morphology parameters
are added to the final catalog ({\tt .mag} file).

It is possible to skip the ALLFRAME stage for certain fields by specifying them in the {\tt alfexclude} option of the {\tt photred.setup} file.  This
option was used for the short SMASH exposures.

\subsubsection{APCOR}
The DAOPHOT program DAOGROW \citep{Stetson90} is used to produce growth-curves for each band and night separately.  These are used to
produce ``total" photometry (including the broad wings) for the bright PSF stars for each chip.  These values are then compared to the PSF photometry
values for the PSF stars (from neighbor subtracted images) produced in the DAOPHOT stage to compute an average aperture correction for each chip.
These values are all stored in the {\tt apcor.lst} file and used later in the CALIB stage.

\subsubsection{ASTROM}
The WCS in the FITS header is used to add the $\alpha$ and $\delta$ coordinates for each object to the catalog.

\subsubsection{CALIB}
The photometry is calibrated using the transformation equations given in the transformation file specified in {\tt photred.setup} (e.g., {\tt n1.trans}).
The equations in the file can pertain to various levels of specificity: (1) only the band is specified, (2) the band and chip are specified, or (3) the band, chip and night are
specified.  The terms in the transformation file are zero-point, extinction, color, extinction$\times$color, and color$\times$color, along with their uncertainties.
Besides these corrections the photometry is also corrected for the exposure time and the aperture correction (for that chip).

Because the {\em calibrated} color is used for the color term,
the software uses an iterative method to calibrate the photometry (using an initial color of zero).  A weighted average value from all exposures is used for the
magnitude in the other band to construct the color (if multiple exposures in that band were taken) but not for the band being calibrated (the value for that exposure is used).
Also, a color of zero is used for objects for which a good color cannot be constructed.  The loop to derive the true source color continues until convergence (all magnitude differences are
below the 0.1 mmag level or 50 iterations, whichever is first).

Calibrated photometry for each exposure (e.g., {\tt G2}, {\tt Z4}) is given in the output file and, optionally,
the average magnitudes per band (e.g., {\tt GMAG}, {\tt ZMAG}) and the instrumental magnitudes for each exposure (e.g., {\tt I\_G2}, {\tt I\_Z4}).
Because for SMASH a global calibration strategy was adopted, all of the values in the transformation file were set to zero so that the photometry was only
corrected for the exposure time and aperture corrections.   

\subsubsection{COMBINE}
The individual chip catalogs are combined to create one catalog for the entire field.  Sources detected in multiple chips (from dithered exposures) are combined
and their photometry combined.  The default matchup radius is 0.5\arcsec~($\sim$2 pixels).

\subsubsection{DEREDDEN}
\citet[][SFD]{SFD98} $E(B-V)$ extinction values for each source are added as a separate column in the final, combined catalog.  Extinction ($A[X]$) and reddening
($E[X-Y]$) values for the bands and colors specified in the {\tt photred.setup} setup file (using $A[X]/E[B-V]$ values from the given {\tt extinction} file) are also added
to the catalog.  The $ugriz$ reddening coefficients from \citet{Schlafly11} were used for SMASH.

\subsubsection{SAVE}
The final ASCII catalog is renamed to the name of the field (e.g., {\tt F5} is renamed to {\tt Field62}) and a copy is created in the IDL ``save" and FITS binary table formats.
In addition, a useful summary file is produced with information on each exposure and chip for that field.

\subsubsection{HTML}
This stage creates static HTML pages to help with quality assurance of the PHOTRED results.  Quality assurance metrics are computed and plots created
for the pages.  This stage was skipped for SMASH since custom quality assurance routines were written.



\subsection{0.9-m Photometry}
We performed photometry on the 0.9-m observations of SDSS standards and SMASH target fields with a pipeline based on the DAOPHOT software suite (by K.O., separate from PHOTRED/STDRED).  In short, we used DAOPHOT to measure aperture-based photometry of the standard star frames, with a smallest aperture of 6\arcsec~diameter ($\sim$15 pixels) and a largest aperture of 15\arcsec~diameter ($\sim$38 pixels).  We used DAOGROW to measure the growth curve based on the aperture measurements and to extrapolate the total instrumental magnitudes of the standard stars.  These total instrumental magnitudes were then used as input to compare with the standard magnitudes in our derivation of the photometric transformation equations from the standard star observations, described in full below.  We also measured PSF photometry of the SMASH target fields and the standards using DAOPHOT and ALLSTAR.  We derived PSFs from the images using as many as 200 point sources per image, using an iterative method to remove neighbors from the PSF stars and to improve the PSF estimation.  We applied aperture corrections to the PSF photometry by comparing the PSF meaurements with the total instrumental magnitudes from DAOGROW and fitting for the residuals with second-order polynomial functions in $x$ and $y$.  These aperture-corrected instrumental PSF magnitudes were used as the input when we derived standard magnitudes for the SMASH fields, described below.  We also measured instrumental PSF magnitudes for the standard fields to make sure that the PSF photometry procedure did not introduce any systematic errors, also described below.

\section{Calibration}
\label{sec:calibration}

The southern sky that SMASH is observing has not been well covered with $ugriz$ CCD imaging, which means that
it is not possible to calibrate our photometry with existing catalogs (in the same area of the sky) as can be done in the north by use of SDSS and
Pan-STARRS1 \citep{Chambers16} data.  Therefore, we must use the traditional techniques of calibrating our data with observations of standard star fields
(on photometric nights) and extra calibration exposures (for non-photometric nights). Our overall calibration philosophy is as follows:
\begin{itemize}
\item For photometric DECam nights calibrate the data using photometric transformation equations derived using DECam exposures of SDSS
equatorial fields.
\item For non-photometric DECam data set the zero-points using photometric DECam data of the same field or overlapping neighboring fields
(only available in the LMC/SMC main bodies).
\item For SMASH fields with no photometric DECam data the calibrated 0.9m photometry for the central region is used to set the zero-points.
\end{itemize}

It is important to note that SMASH is on a quasi-SDSS photometric system.  The data were obtained through the DECam passbands and
calibrated onto the SDSS system using zero-points and {\em linear} color-terms.  However, the data were {\em not} corrected for non-linear effects that arise because
of the differences in the SDSS and DECam passbands, which are significant for the $r$-band and $u$-band.  Therefore, the SMASH colors of stars can deviate
from how they would appear in SDSS in certain bands for very red or blue stars.  Caution should be used when comparing the
data to other catalogs or model isochrones in these situations.

\subsection{Standard Star Calibration of DECam Data}
\label{subsec:decamcalib}

On every DECam night ``standard star" observations in the $ugriz$ bands were taken every couple of hours of SDSS equatorial fields at both high and low airmass.
The equatorial fields lie in Stripe 82 and Stripe 10, and are the same as those chosen by DES to regularly sample the full range in RA.
We downloaded ``reference" catalogs via CasJobs\footnote{\url{http://casjobs.sdss.org}} for the equatorial fields from the SDSS DR12 \citep{DR12}.
These observations generally provided several thousands of ``standard star" measurements per exposure and good color coverage.

\subsubsection{Processing of Standard Star Data with STDRED}
\label{subsubsec:stdred}

To reduce the DECam standard star exposures we use the STDRED pipeline, which is a sister package to PHOTRED and works in a similar
manner.  The same {\tt SMASHRED\_PREP.PRO} pre-processing script is used to uncompress, mask and split the CP-reduced images and download
the astrometric reference catalogs per field.  The main STDRED steps that are used by SMASH are:

\begin{itemize}
\item WCS: Fits the chip WCS using the astrometric reference catalog.
\item APERPHOT: Detects sources and performs aperture photometry.
\item DAOGROW: Calculates aperture corrections via curves of growth and applies them to the aperture photometry.
\item ASTROM: Adds $\alpha$/$\delta$ coordinates to the photometric catalog.
\item MATCHCAT: Cross-matches the observed catalog with the reference catalog and for matches outputs merged information from both catalogs.
\item COMBINECAT: Combines all of the matched photometry for a given filter.
\item FITDATA: Fits photometric transformation equations (with zero-point, color, and extinction terms) for each filter using all of the data.  However,
new custom software was developed to derive these equations for SMASH (see next section).
\end{itemize}

The standard star exposures from each DECam run are processed with STDRED in their own directory.

\subsubsection{Derivation of DECam Photometric Transformation Equations from Standard Star Data}
\label{calib:transphot}

To produce the highest-quality and most uniform calibration we decided to write new custom software to determine the DECam
photometric transformation equations using all of the standard star data together.  Transformation equations of the following form were used:

\noindent
\begin{math}
m_{\rm obs} = m_{\rm cal} + {\rm ZPTERM} + {\rm COLTERM}\times(color_{\rm cal}) + {\rm AMTERM}\times X
\end{math}

\noindent
where ZPTERM is the zero-point term, COLTERM is the color term, $color_{\rm cal}$ is the calibrated color (which includes the band being calibrated),
AMTERM is the extinction term, and $X$ is the airmass.

\begin{figure}[t]
\includegraphics[width=1.0\hsize,angle=0]{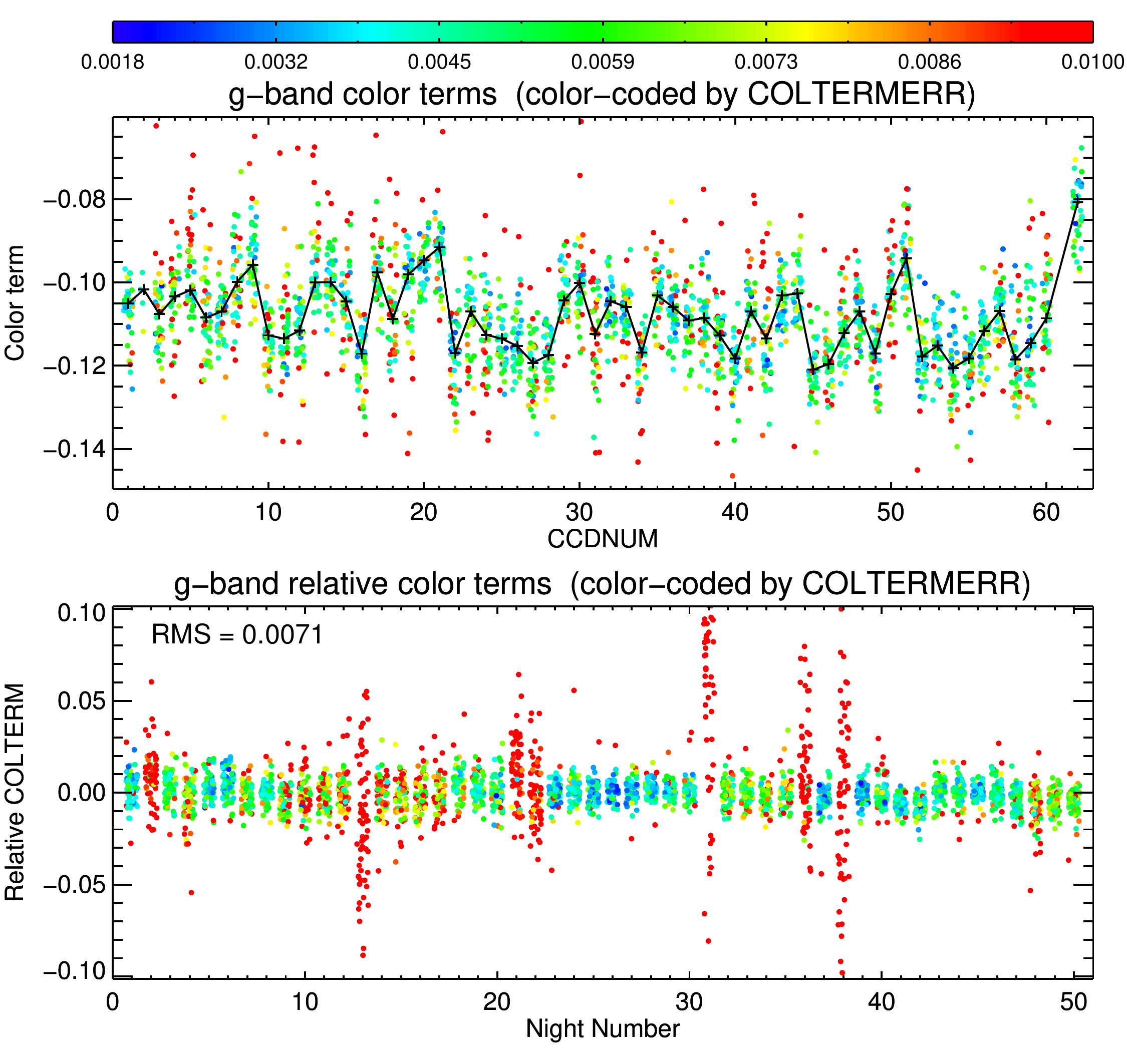}
\caption{The $g$-band color terms of the DECam photometric transformation equations.
(Top) The dependence of the color terms on chip number.  (Bottom) The temporal dependence of the 
``relative" color terms with the median color term of each chip subtracted (``night number" is a running counter of SMASH nights).
The nights with large scatter are non-photometric nights.
}
\label{fig_colterm}
\end{figure}

The new software  ({\tt SOLVE\_TRANSPHOT.PRO}) has several options for what variables to fit or hold fixed (zero-point, color, extinction, and color$\times$extinction terms) 
and over what dimensions (e.g., night and chip) to average or ``bin" values.  At first all variables (zero-point, color and extinction terms) were fit separately
for each night and chip combinations to see how much the terms vary and over what dimensions.

\begin{center}
\begin{deluxetable*}{ccrrr}
\tablecaption{SMASH Median Photometric Transformation Equations}
\tablecolumns{5}
\tablewidth{300pt}
\tablehead{
\colhead{Band} & \colhead{Color} & \colhead{Zero-point term} & \colhead{Color term} & \colhead{Extinction term}
} 
\startdata
$u$  &  $u-g$ & 1.54326 $\pm$   0.0069 &     0.0142 $\pm$   0.0041  &     0.3985 $\pm$   0.00240 \\
$g$  & $g-r$ & $-$0.3348 $\pm$   0.0019 &     $-$0.1085 $\pm$   0.0010  &     0.1747 $\pm$  0.00076  \\
$r$   & $g-r$ & $-$0.4615 $\pm$   0.0018 &    $-$0.0798 $\pm$   0.0011  &    0.0850 $\pm$  0.00098  \\
$i$   &  $i-z$ & $-$0.3471 $\pm$   0.0016 &     $-$0.2967 $\pm$   0.0012  &    0.0502 $\pm$  0.00058  \\
$z$  &  $i-z$ & $-$0.0483 $\pm$   0.0023 &    $-$0.0666 $\pm$   0.0016  &    0.0641 $\pm$  0.00075 \\
\enddata
\label{table_transphot}
\end{deluxetable*}
\end{center}

{\bf Color:}  We found that the color terms vary from chip to chip (at the $\sim$0.01 mag level as also noted on the DECam Calibration
webpages\footnote{\url{http://www.ctio.noao.edu/noao/node/3176}}; see the left-hand panel of Figure \ref{fig_colterm}),
but they appear to be temporally stable (see the bottom panel of Figure \ref{fig_colterm}).  Therefore, we fit the (linear) color terms for each chip separately
by taking a robust average over all photometric nights.

\begin{figure}[b]
\begin{center}
\includegraphics[width=1.0\hsize,angle=0]{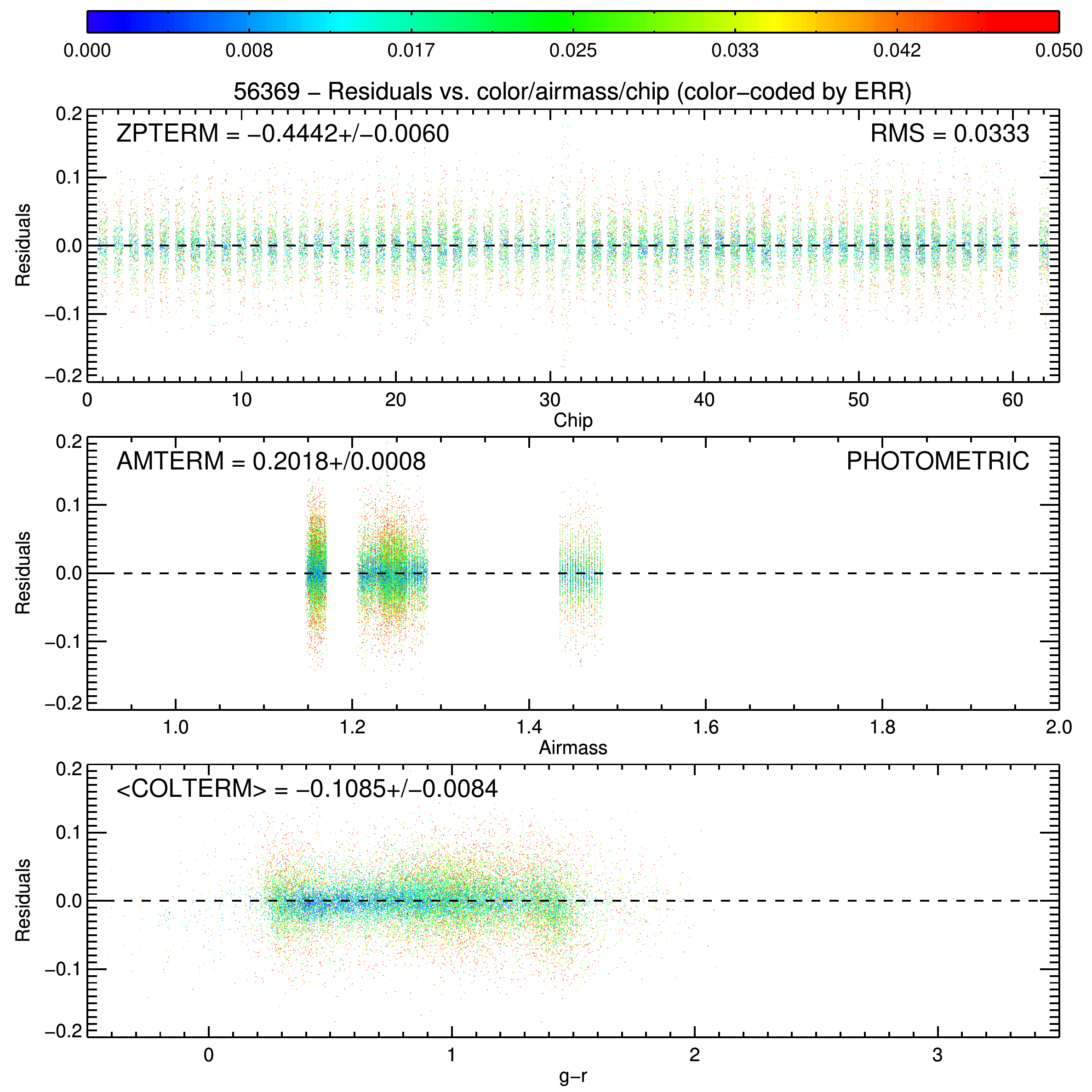}
\end{center}
\caption{The $g$-band residuals (SDSS reference magnitude $-$ derived magnitude using the transformation equations) after fitting the
photometric transformation equations to standard star observations for a typical photometric night.
The observations are color-coded by their photometric error. (Top)  Residuals versus chip number.  (Middle) Residuals versus airmass.
(Bottom) Residuals versus $g$$-$$r$ color. The derived terms of the photometric transformation equations and their uncertainties are given
in the upper left-hand corner of their respective panel.}
\label{fig_stdresid}
\end{figure}

No evidence for systematics was found in the color residuals of $g$/$i$/$z$ indicating there was no need for higher order color terms.  
For $u$-band there are systematics in the residuals (consistent across all fields) that would require higher order terms to fit.  This is largely because
of the different throughput curves for the SDSS and DECam filters.  We decided not to add higher order terms as these could adversely affect very blue
or red objects (where the solution is not well constrained).  However, to determine a uniform and reliable zero-point we decided to fit the shape in the residuals
and remove this pattern from the observed data at the very beginning of the procedure.  In addition, we restricted the color range to 1.0 $<$ $u$-$g$ $<$ 2.5.
After this correction and color restriction are applied the residuals are flat.  We similarly restrict the color range for $r$ band ($g$-$r$ $<$ 1.2) because
the correlation between SDSS and DECam $r$-band magnitudes becomes non-linear for redder stars due to the difference in the throughput curves.

{\bf Extinction:} An appreciable number of nights had a small range in airmass for the standard star observations that produced unreliable
extinction term measurements.  Therefore, for these nights we calculated a weighted (by uncertainty and time difference) average of the nightly extinction
terms for the four closest neighboring good nights.  Similarly, for nights with larger airmass ranges, we improve the accuracy by
refitting the extinction term using the individual data points from the four closest good neighboring nights (but these must be within 30 days).
Finally, we found that there was no appreciable color$\times$extinction dependence and, therefore, these terms were not 
included in the fits.

{\bf Zero-point:} We tried separating the zero-points into nightly zero-points and relative chip-to-chip (for each band but constant with time) zero-point offsets,
considering that although the zero-point can change nightly, due to transparency and extinction variations, the zero-points of one
chip to another (in a given band) should remain the same.  We found, however, that the scatter in the relative chip-dependent zero-points over the many
nights was somewhat higher than was anticipated (but still small at $\sim$0.01 mag) and we obtained better results by fitting a zero-point for each night and chip combination.
Therefore, we adopted the latter strategy and ``abandoned" the relative zero-points (although they are computed and saved in the final output file).

Photometric nights are determined by seeing if the observers noticed any sign of clouds, looking for cloud cover in the CTIO RASICAM all-sky infrared
videos\footnote{\url{http://www.ctio.noao.edu/noao/node/2253}},
and, finally, by looking at the scatter in the standard star residuals.
The full list of nights for which STDRED was run and the photometric status are given
in {\tt smash\_observing\_conditions.txt} (in {\tt SMASHRED/obslog/}).

\begin{center}
\begin{deluxetable*}{lrrrrrrr}
\tablecaption{0.9-m Photometric Transformation Equations (SDSS filter set)}
\tablecolumns{8}
\tablewidth{400pt}
\tablehead{
 & \colhead{Scale factor} & \colhead{Extinction} &\colhead{$x$ factor} & \colhead{$y$ factor}&\colhead{Time factor} & \colhead{Color term} & \colhead{Zero point} \\
\colhead{Band} & \colhead{$(ABCDE)_1$} & \colhead{$(ABCDE)_2$} &\colhead{$(ABCDE)_3$} & \colhead{$(ABCDE)_4$}&\colhead{$(ABCDE)_5$} & \colhead{$(ABCDE)_6$} & \colhead{$(ABCDE)_7$} \\
}
\startdata
\sidehead{14 -- 23 Feb 2014}
$u$ & 1.001 $\pm$ 0.002  & 0.51 $\pm$ 0.02 & $\equiv$ 0.0 & $\equiv$ 0.0 & $\equiv$ 0.0 & $-$0.034  $\pm$ 0.004 & 4.59 $\pm$ 0.04 \\
$g$ & 0.997 $\pm$ 0.001  & 0.19 $\pm$ 0.01 & 0.0 & 0.0 & 0.0 & 0.009  $\pm$ 0.011 & 2.66 $\pm$ 0.03 \\
$r$ & 0.995  $\pm$ 0.001 & 0.11 $\pm$ 0.01 & 0.0 & 0.0 & 0.0 &$-$0.022  $\pm$ 0.007 & 2.67 $\pm$ 0.02 \\
$i$ & 0.995 $\pm$ 0.002  & 0.06 $\pm$ 0.01 & 0.0 & 0.0 & 0.0 &$-$0.017  $\pm$ 0.014 & 3.13 $\pm$ 0.03 \\
$z$ & 0.998 $\pm$ 0.001  & 0.07 $\pm$ 0.02 & 0.0 & 0.0 & 0.0 & 0.040  $\pm$ 0.011 & 3.95 $\pm$ 0.02 \\
\sidehead{25 Sep -- 2 Oct 2014, 26 Apr -- 3 May 2015, and 27 -- 29 Nov 2015}
$u$ & $\equiv$ 1.0 & 0.49 $\pm$ 0.02 & $\equiv$ 0.0 & $\equiv$ 0.0 & $\equiv$ 0.0 & $-$0.034 $\pm$ 0.004  & 4.17 $\pm$ 0.29 \\
$g$ & 1.0 & 0.18 $\pm$ 0.01 & 0.0 & 0.0 & 0.0 & 0.005 $\pm$ 0.011 & 2.51 $\pm$ 0.31 \\
$r$ & 1.0 & 0.10 $\pm$ 0.01 & 0.0 & 0.0 & 0.0 & $-$0.028  $\pm$ 0.007 & 2.49 $\pm$ 0.21 \\
$i$ & 1.0 & 0.06 $\pm$ 0.01 & 0.0 & 0.0 & 0.0 & $-$0.026  $\pm$ 0.014 & 2.91 $\pm$ 0.12 \\
$z$ & 1.0 & 0.06 $\pm$ 0.01 & 0.0 & 0.0 & 0.0 & 0.022  $\pm$ 0.011 & 3.72 $\pm$ 0.05 \\
\enddata
\label{09m_trans}
\end{deluxetable*}
\end{center}

The $\sim$3100 variables (zero-points for $\sim$50 nights $\times$ 60 chips, color terms for 60 chips, extinction terms for $\sim$50 nights) were not fit to the data simultaneously but were found
through an iterative fitting process:
\begin{enumerate}
\item Fit all terms separately for all night and chip combinations.
\item Compute the mean color term per chip using only photometric data.
\item Fix color terms and refit zero-point and extinction terms.
\item Average extinction terms.  For photometric nights with poor solutions or low airmass ranges a weighted average of the extinction terms of the
nearest four neighboring photometric nights is computed.  For the rest of the photometric nights, a new extinction term is computed using data included
from the four nearest neighboring nights.  The extinction term is set to zero for non-photometric nights.
\item Fix color and extinction terms and refit zero-point terms.
\end{enumerate}
While solutions are found for all nights, only the transformation equations for photometric nights are used to calibrate the data.

The final photometric transformation equations are written to file ({\tt smashred\_transphot\_eqns.fits} available in {\tt SMASHRED/data/}) with zero-point, color and
extinction terms (with uncertainties and averaging information) for each night and chip combination, as well as separate tables with information unique to each chip
(e.g., color term) and information unique to each night (e.g., extinction term).  The formal uncertainties on the terms are:
$\sim$0.002, $\sim$0.0015, and $\sim$0.0007 for the zero-point, color and extinction, respectively.
For an average color and airmass this amounts to a formal uncertainty in the photometry of $\sim$0.002 mag (0.009 mag for $u$).
Table \ref{table_transphot} gives median values and uncertainties per band, while example residuals versus chip, airmass and color for a single night are shown in
Figure \ref{fig_stdresid}.

The nights of the UT 2014 January 5--7 observing run were clear and photometric but no SDSS standard star observations were taken.  Therefore, the regular procedures
could not be used to determine the transformation equations for these nights.  Subsequently some of the fields from this run could be calibrated because they were
reobserved on other photometric nights (with standard star data) or 0.9-m calibration data were obtained.  The photometric transformation equations were then determined
(``backed-out") by using these calibrated fields and using the previously derived chip-dependent color terms.  The {\tt smashred\_transphot\_eqns.fits} was then updated
with these values and the data for those nights could be calibrated in the regular manner.

\subsection{Calibration of 0.9-m Data}
Using the aperture corrected 0.9-m photometry, we explored fits to equations of the form:

\noindent
\begin{math}
u_{\rm obs} = A_1u + A_2X + A_3x + A_4y + A_5t + A_6(u-g) + A_7 \\
g_{\rm obs} = B_1g + B_2X + B_3x + B_4y + B_5t + B_6(g-r) + B_7 \\
r_{\rm obs} = C_1r + C_2X + C_3x + C_4y + C_5t + C_6(g-r) + C_7 \\
i_{\rm obs} = D_1i + D_2X + D_3x + D_4y + D_5t + D_6(r-i) + D_7 \\
z_{\rm obs} = E_1z + E_2X + E_3x + E_4y + E_5t + E_6(i-z) + E_7,
\end{math}

\noindent
where $u_{\rm obs}g_{\rm obs}r_{\rm obs}i_{\rm obs}z_{\rm obs}$ are instrumental magnitudes, $ugriz$ are standard SDSS magnitudes drawn from \citet{Smith02} and from SDSS DR12 \citep{DR12}, $X$ is the airmass, $x$ and $y$ are pixel positions on the detector, and $t$ is time of observation during the night.  

We fit this set of equations first to the data taken on the almost entirely photometric run from 14 -- 23 Feb 2014.  Table \ref{09m_trans} shows the best-fit coefficients.  While we fit the transformation equations independently on each of the ten nights, we show only the average coefficients and their standard deviations in the table, as values were in all cases consistent across the nights.  From our fits, we found no evidence for strong pixel position-dependent or time-dependent terms.  We did, however, find evidence for a small magnitude-dependent scale factor of 0.1--0.5\%, which may point to a small non-linearity with the Tek2K CCD.

\begin{figure}[t]
\begin{center}
\includegraphics[width=1.0\hsize,angle=0]{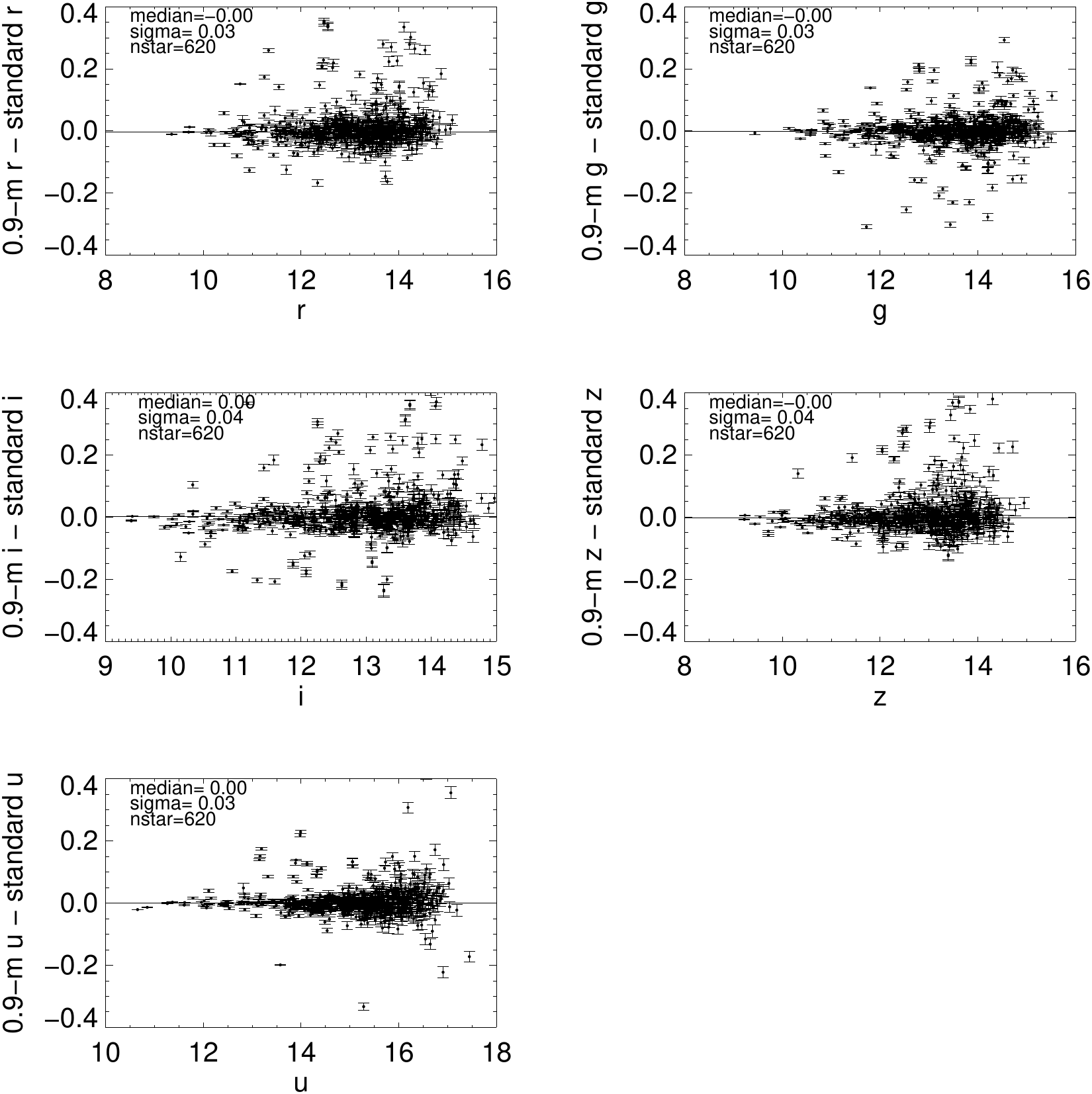}
\end{center}
\caption{Residuals of the 0.9-m photometry relative to the standard star data versus magnitude for the $ugriz$ bands.
Statistics for the residuals are in the upper left-hand corner indicating rms values of 0.03--0.04 mag.}
\label{fig_09mresid}
\end{figure}

We next explored fits to the equations for the observing runs 25 Sep -- 2 Oct 2014, 26 Apr -- 3 May 2015, and 27 -- 29 Nov 2015.  These runs were complicated by variable weather conditions, work on the camera electronics that changed the gain setting of the CCD, and by a temporary change from the CTIO SDSS $griz$ filter set to the PreCam $griz$ filter set that more closely matches the filter set used by DECam.  For these observations, we fit the SDSS and PreCam sets separately.  For each filter set, we first used all of the nights observed with that set to measure the color term coefficients. To do this, we allowed the zero point for each frame to be fit independently, which removes all other variables from consideration other than the color term; this allowed us to use standards taken on non-photometric nights to constrain the color term.  We then fixed the color term coefficients to these fitted values, and for the photometric nights fit for the remaining coefficients on a per night basis.  For these fits, we found no evidence for pixel position-dependent, time-dependent terms, or, in contrast to the Feb 2014 observations, a magnitude-dependent scale term.  Table \ref{09m_trans} shows the fitted coefficients for the SDSS filters, where the larger standard deviation in the tabulated zero point reflects the large span of time over which the observations were taken.
Figure \ref{fig_09mresid} shows the photometric residuals for the standard star fields.  Table \ref{09m_trans_des} shows the color terms for the DES PreCam filter set, for which we did not derive full transformation equations because of weather that was not completely photometric.  We found some significant differences between the DES PreCam and SDSS color terms, particularly for the $r$, $i$, and $z$ filters, as is to be expected from the differences in the bandpasses.  The color terms are not, however, identical to those measured with DECam, which we ascribe to differences in the PreCam bandpasses compared to DECam, as well as to differences in the telescope and detector response functions.  In the end, we did not use the PreCam data for calibration, but present the results to verify that using a filter set that is a closer match to those used with DECam reveals systematic differences with the SDSS filters.

\begin{center}
\begin{deluxetable}{lr}
\tablecaption{0.9-m Photometric Color Terms (DES PreCam filter set)}
\tablecolumns{2}
\tablewidth{150pt}
\tablehead{
 & \colhead{Color term} \\
\colhead{Band} & \colhead{$(ABCDE)_6$} \\
}
\startdata
\sidehead{30 Apr -- 3 May 2015}
$u$ & $-$0.033$\pm$0.01   \\
$g$ &  0.020$\pm$0.001   \\
$r$ & $-$0.068$\pm$0.001   \\
$i$ & $-$0.063$\pm$0.001   \\
$z$ & $-$0.026$\pm$0.002   \\
\enddata
\label{09m_trans_des}
\end{deluxetable}
\end{center}

\subsection{Calibration Software}
\label{subsec:calibsoftware}

New software was developed to perform calibration of SMASH fields across multiple nights and using a variety of zero-point calibration
methods.  The software also takes advantage of the overlap of our multiple short exposures with large dithers to tie all of the chip data for a given field
onto the same photometric zero-point using an \"ubercal technique.
While PHOTRED performs similar tasks (i.e., COMBINE and CALIB), it is on a night-by-night basis.  This meant, therefore,
that the custom SMASH calibration software needed to start with the instrumental PHOTRED photometry catalogs output by the ASTROM stage. 

The calibration follows these steps (in pseudocode): \\

WHILE calibrated photometry changes $>$1 mmag:
\begin{itemize}
\item The photometry is calibrated using the zero-point (ZPTERM), color (COLTERM) and extinction/airmass terms (AMTERM).  This is an iterative process
because of the color-term.  First, the source photometry is calibrated using the average photometry to construct the color (a color of zero is used on the first iteration),
then the photometry is averaged per object and band.  The process repeats until all changes are $<$0.1 mmag.
  ({\tt SMASH\_APPLY\_PHOTTRANSEQN.PRO}) \\
\item[] FOR all filters
\begin{enumerate}
\item Measure the pair-wise photometric offsets of overlapping chips  ({\tt SMASH\_MEASURE\_MAGOFFSET.PRO}).
\item Solve the relative magnitude offsets per chip using the \"ubercal algorithm  ({\tt SMASH\_SOLVE\_UBERCAL.PRO}).
\item Determine the photometric zero-point  ({\tt SMASH\_SET\_ZEROPOINTS.PRO}).
\end{enumerate}
\item[] ENDFOR
\end{itemize}
ENDWHILE \\

\begin{figure*}[t]
\begin{center}
$\begin{array}{ccc}
\includegraphics[width=0.33\hsize,angle=0]{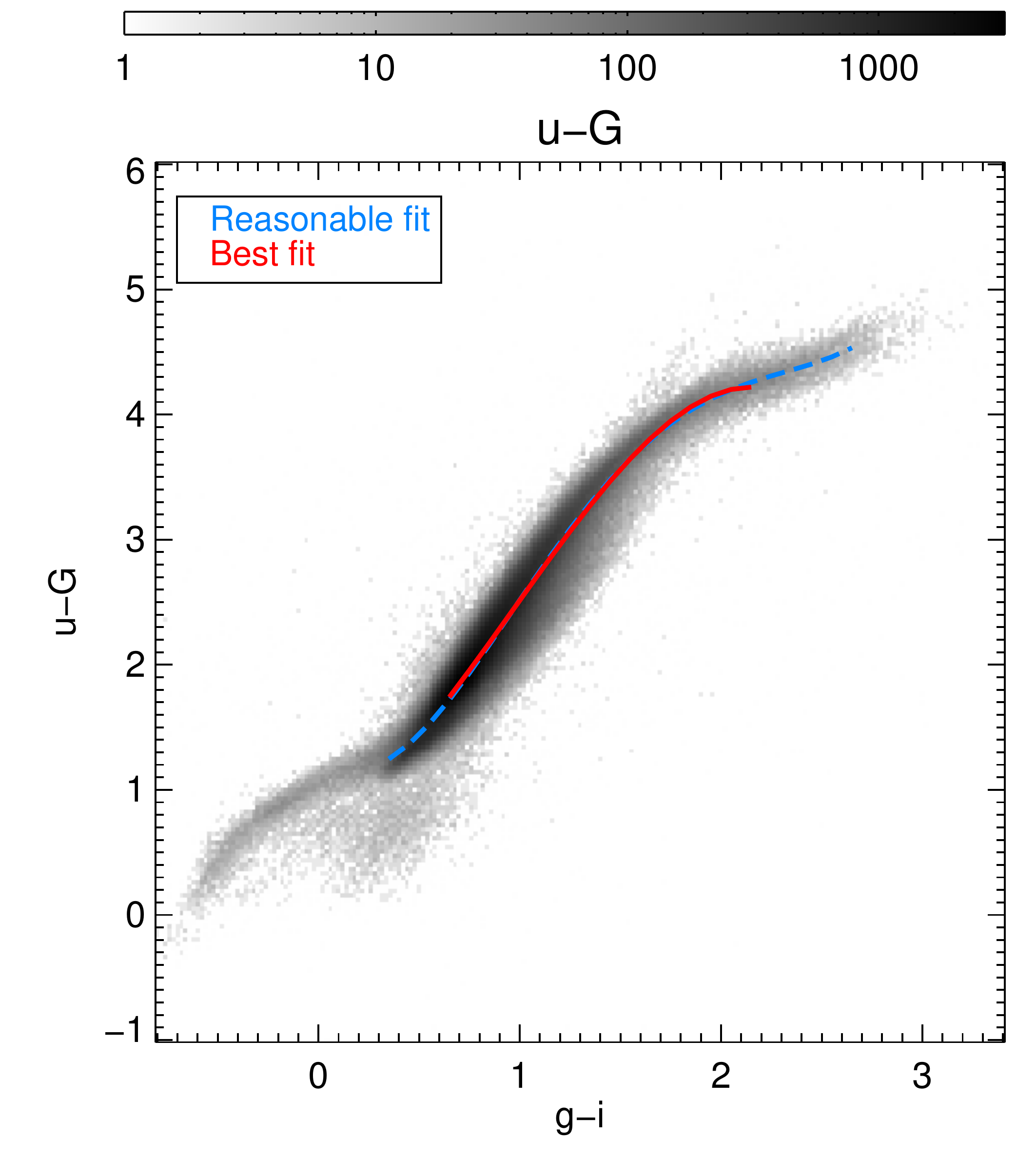}
\includegraphics[width=0.33\hsize,angle=0]{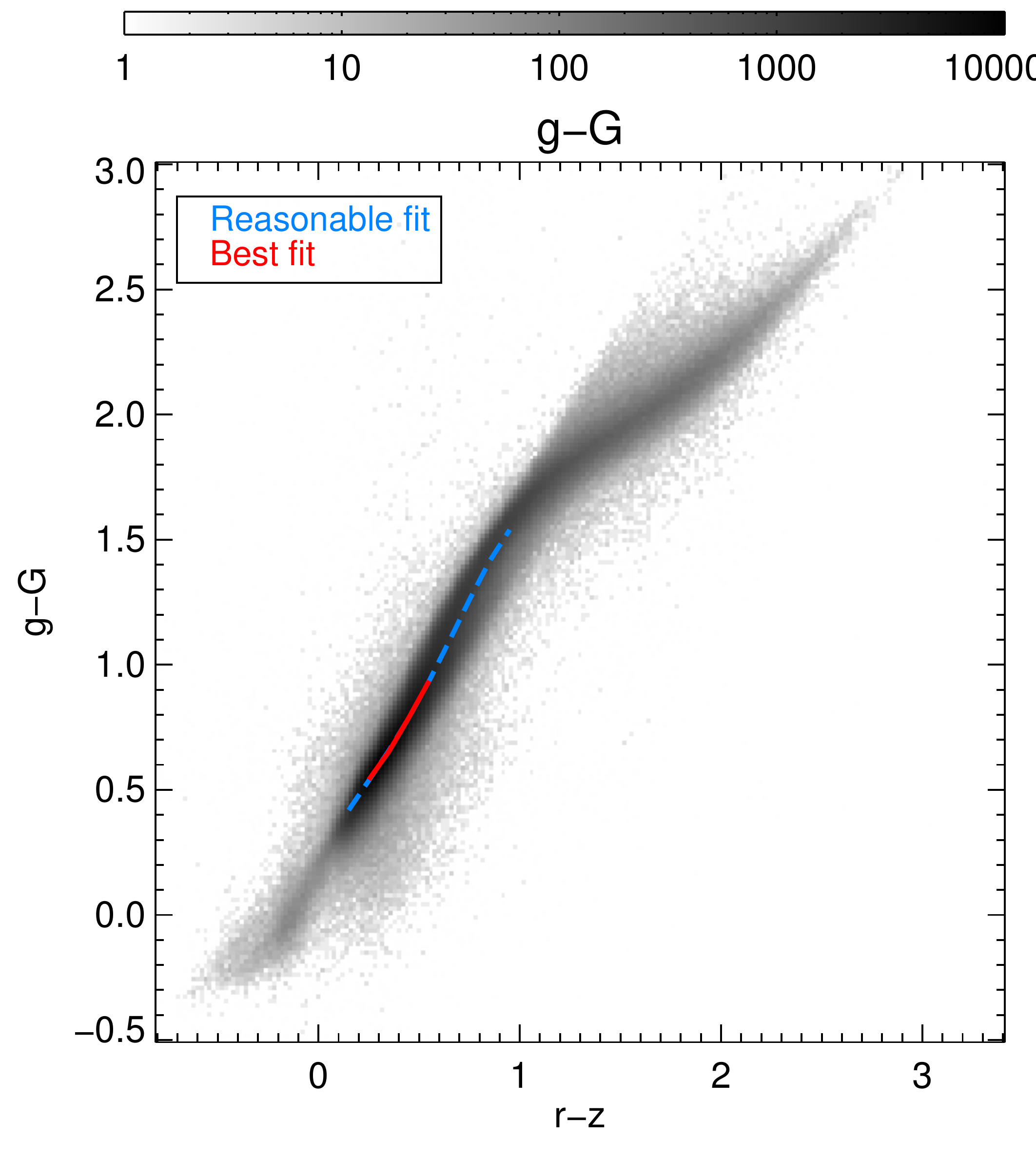}
\includegraphics[width=0.33\hsize,angle=0]{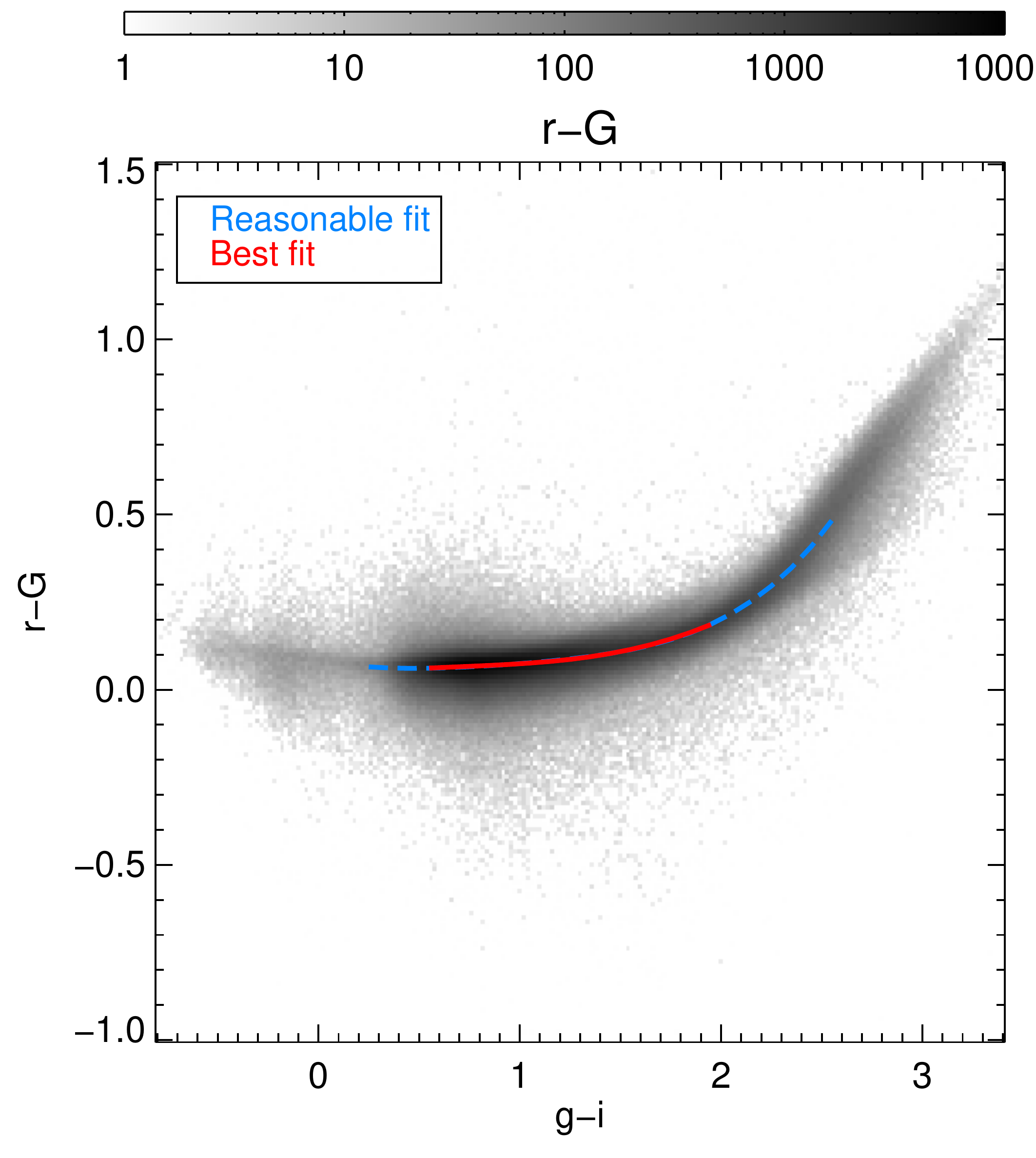}  \\
\includegraphics[width=0.33\hsize,angle=0]{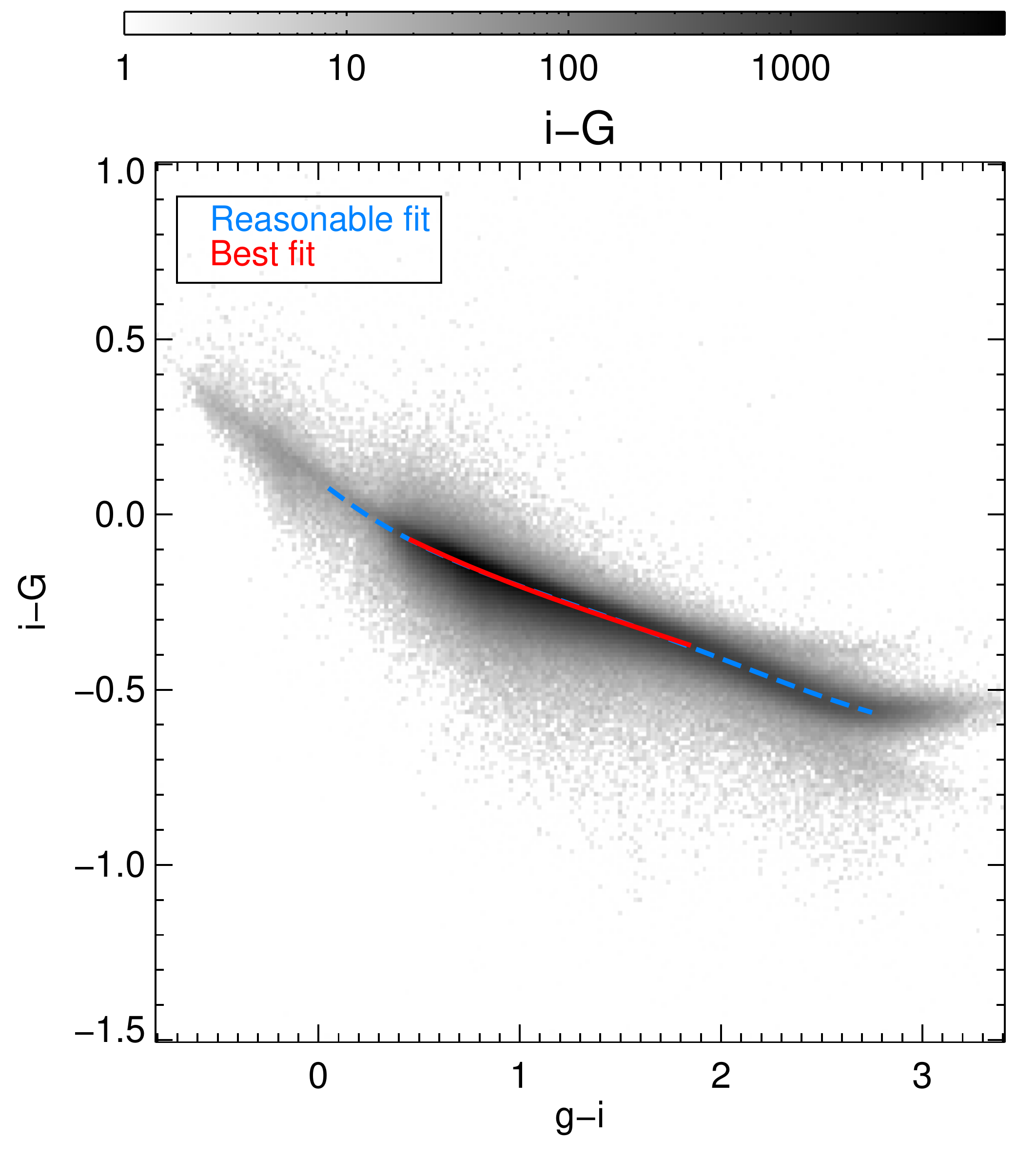}
\includegraphics[width=0.33\hsize,angle=0]{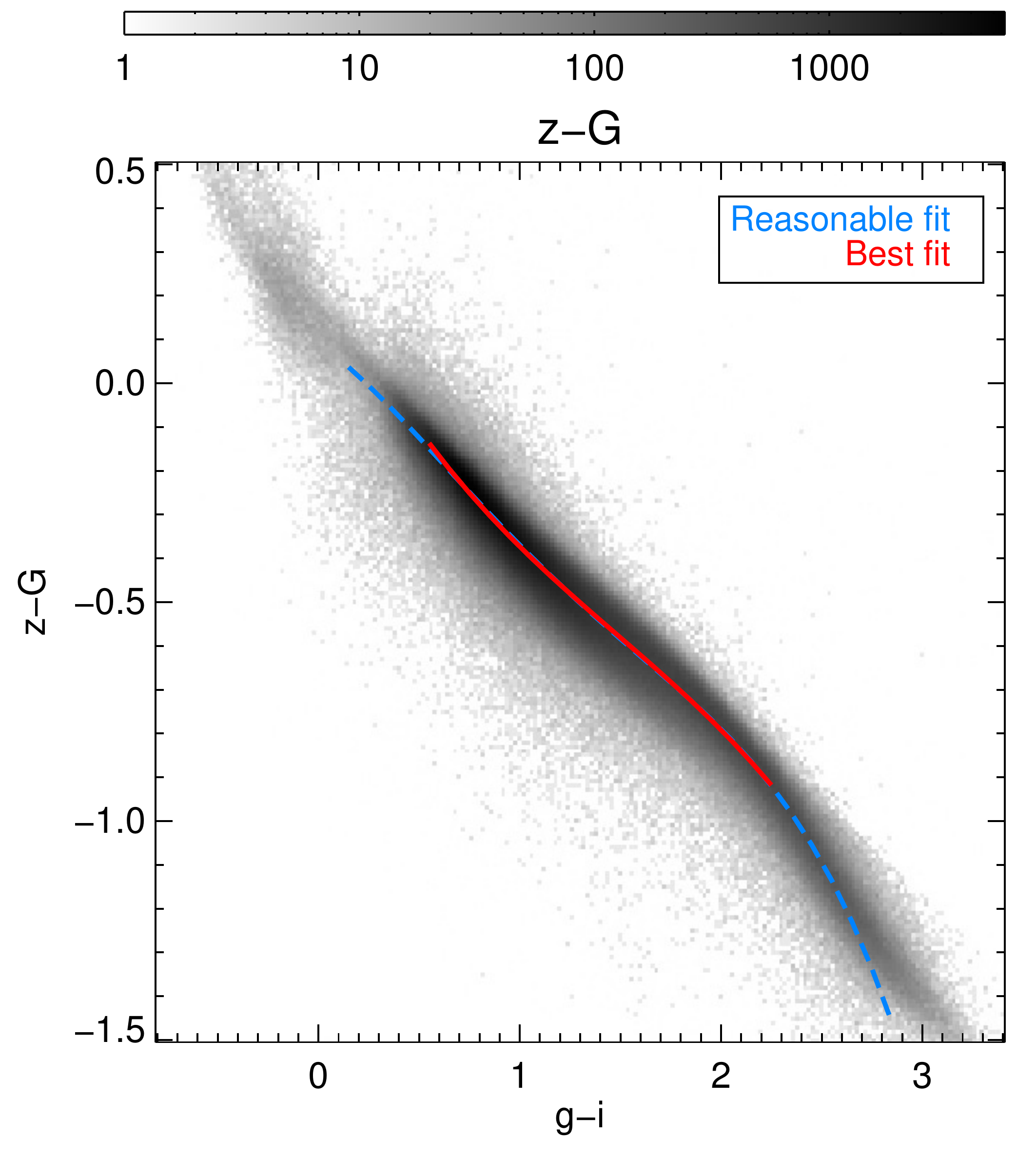}
\end{array}$
\end{center}
\caption{SMASH-Gaia color-color distributions and relations.  Blue dashed lines are polynomial fits over a ``reasonable" color range (avoiding the highest rms regions at the ends)
while the solid red lines are polynomial fits over the ``best" color range giving the lowest rms scatter.   The scatters over the best color ranges are
6\% ($u$), 1\% ($g$), 0.2\% ($r$), 0.4\% ($i$) and 0.5\% ($z$).}
\label{fig_smashgaia}
\end{figure*}

We employ a simple iterative \"ubercal solving technique \citep{P08}.  After all of the pair-wise photometric offsets of overlapping chips are measured, the robust
weighted average offset of a chip relative to its overlapping neighbors is calculated and {one half} of this is used as the \"ubercal correction for this chip. 
The pair-wise photometric offsets are updated for these chip-wise corrections and the procedure repeats until convergence is reached (the average
relative offset change from one iteration to the next is less than 1\%).  The changes become very small after only a couple iterations.
The cumulative corrections are applied to the chip-wise zero-point terms ({\tt ZPTERM}) and saved in the {\tt UBERCAL\_MAGOFFSET} columns.
The \"ubercal technique only measures and solves for a constant magnitude offset for every chip.  There is no allowance for spatial variations across
the chip such as due to variable throughput. The outer (while) loop in the calibration is used to make sure the color
terms are properly taken into account.

One of three different techniques is employed to set the photometric zero-point of the data depending on the observing conditions and what 0.9-m
calibration data are available.  The options in decreasing order of preference are:
\begin{enumerate}
\item Photometric DECam data ({\tt ZPCALIBFLAG}=1):  Any DECam data taken during photometric conditions ({\tt PHOTOMETRIC}=1) and having good photometric transformation
equations from standard star exposures ({\tt BADSOLN}=0) are used to set the photometric zero-point.  Any non-photometric data are tied to this via the
\"ubercal offsets.
\item Overlap with photometric DECam data ({\tt ZPCALIBFLAG}=2): A field with no photometric data itself but that overlaps a neighboring field (this happens mainly in the
central LMC and SMC fields) with photometric data can be calibrated using the overlap.  The median offset of bright, high $S/N$ overlap stars is used to set the zero-point. 
\item 0.9-m calibration data ({\tt ZPCALIBFLAG}=3): If a field cannot be calibrated using the first two options and 0.9-m calibration data are available for the field, then it is
used to determine the zero-point with the stars detected in both the DECam and 0.9-m data.
\end{enumerate}
The type of zero-point calibration used for any field can be found in Table \ref{table_smashfields} as well as in the {\tt FIELD\_chips} file (see below) and {\tt chip} table of the SMASH database.

\begin{figure}
\includegraphics[width=0.99\hsize,angle=0]{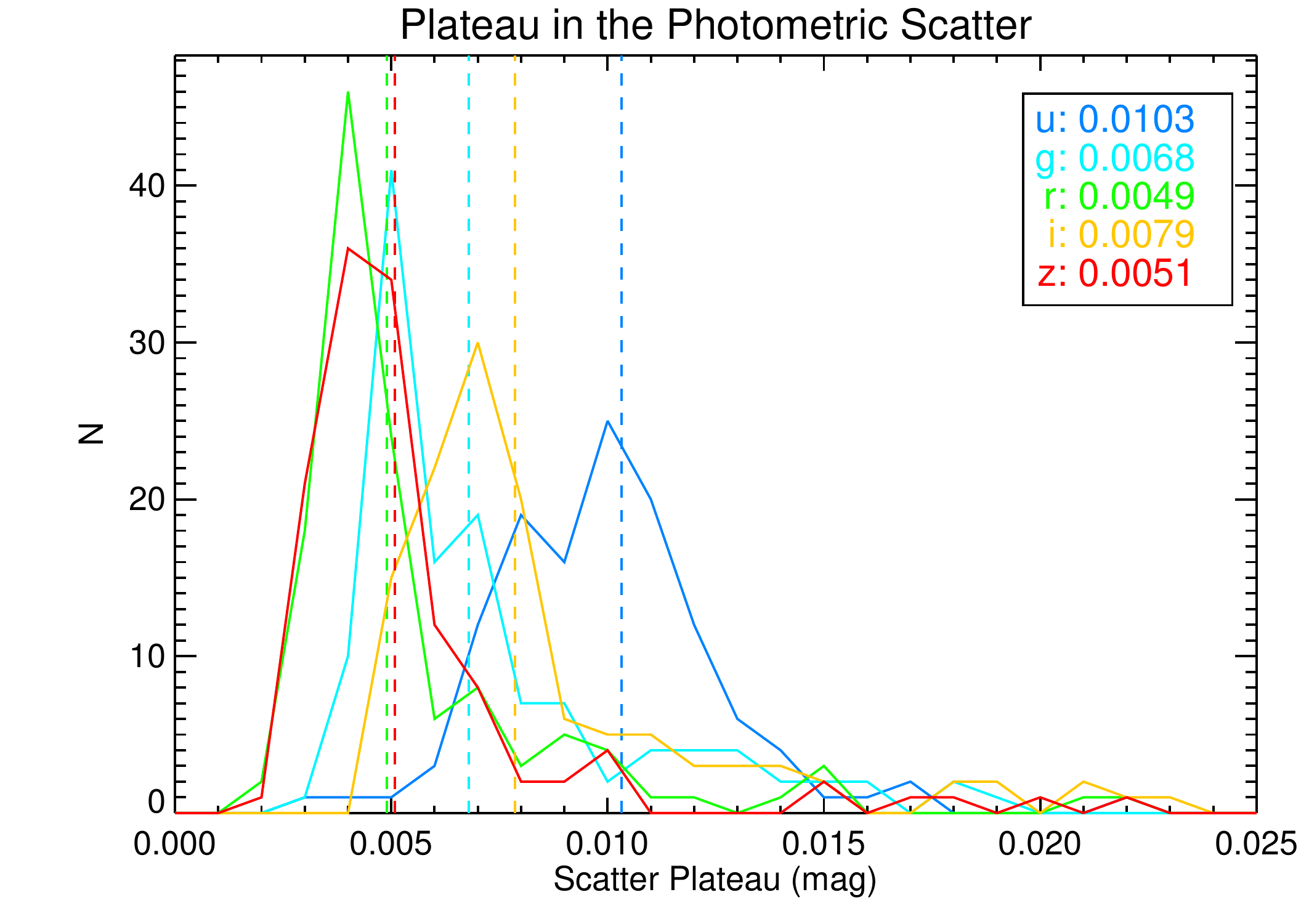}
\caption{The distribution of the lower plateau in the photometric scatter (using multiple measurements of bright stars) in 126 calibrated SMASH fields. This is a good estimate for
the photometric precision of the survey.  Vertical dashed lines show the median value for each band.
}
\label{fig_photscatter}
\end{figure}

\begin{figure*}[t]
\includegraphics[width=1.0\hsize,angle=0]{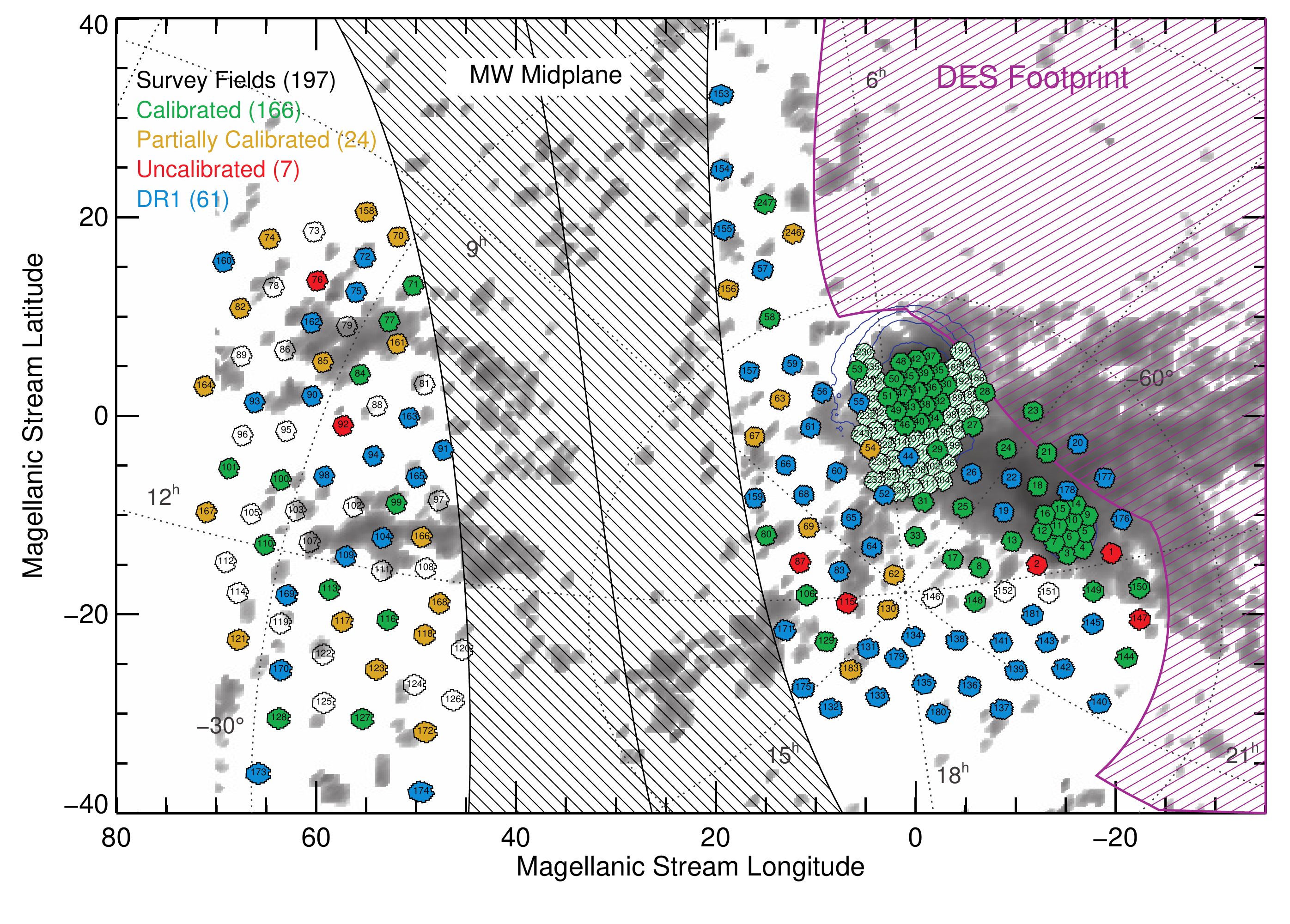}
\caption{The SMASH survey.  The observed \hi column density of the Magellanic Stream system is shown in grayscale \citep{Nidever10}.  Observed SMASH fields are shown as filled hexagons
while unobserved SMASH fields are indicated by open black hexagons.  Green (and dark blue) fields are fully calibrated (166 fields), golden are partially calibrated (some bands calibrated
and some bands uncalibrated; 24 fields), and red are uncalibrated (7 fields).  The green hashed hexagons are the 40 shallow LMC fields.
The 61 DR1 fields are shown in dark blue (all fully calibrated).  The DES footprint is represented by the purple shaded region.
}
\label{fig_dr1map}
\end{figure*}

For fields where none of these options are available we use SMASH-Gaia color-color relations to calculate rough zero-points ({\tt ZPCALIBFLAG}=4).  These relations
were derived by cross-matching 49 of our SMASH fields with good, calibrated photometry and that lie far from both the LMC and SMC against the Gaia catalog.
Bright stars were used to determine the functional relationship between X$_{\rm SMASH}$-G$_{\rm Gaia}$ and a SMASH color ($g$-$i$ for all SMASH
bands except $r$-$z$ for $g$).  These relations are very tight for the redder bands ($r$, $i$, and $z$) with only a scatter of $\sim$0.5\%
(see Figure \ref{fig_smashgaia}), but are poorer and with large color term for the bluer bands ($u$ and $g$ with scatter of $\sim$6\% and $\sim$1\% respectively).
These rough calibrations are a temporary measure.  The remaining partially calibrated or uncalibrated fields will be fully calibrated once the appropriate calibration
data have been obtained.

Once all of the data are calibrated, average coordinates and morphological parameters (e.g., sharp, chi) are computed (weighted averages) from
the multiple measurements of each object.  We then produce an exposure map for the field in each band (at the pixel level) and use this to sort out
non-detections (set to 99.99) from cases of no good data for an object (set to NaN).  \citet{SFD98} E($B$-$V$) extinctions are also added for each object,
but dereddened magnitudes are not computed.
Care should be taken in using the SFD extinction values in the central regions of the MCs because they can be unreliable there and, also, for MW stars
for which they can overestimate the foreground dust.
Finally, the unique objects are cross-matched with the Gaia, 2MASS and ALLWISE catalogs.

\section{Description and Achieved Performance of Final Catalogs}
\label{sec:performance}

The SMASH dataset includes 5,809 DECam exposures with 349,046 separate chip files producing 3,992,314,414 independent source measurements of 418,642,941 unique objects
(296,223,749 with multiple detections).

\subsection{Final Catalog Files}
\label{subsec:catalogdescription}

The final catalogs consist of seven gzip-compressed binary FITS files per field:
\begin{enumerate}
\item {\tt FIELD\_exposures.fits.gz} -- Information about each exposure.
\item {\tt FIELD\_chips.fits.gz} --  Information about each chip.
\item {\tt FIELD\_allsrc.fits.gz} -- All of the individual source measurements for this field.
\item {\tt FIELD\_allobj.fits.gz} -- Average values for each unique object.
\item {\tt FIELD\_allobj\_bright.fits.gz} -- Bright stars from allobj used for cross-matching between fields.
\item {\tt FIELD\_allobj\_xmatch.fits.gz} -- Cross-matches between SMASH and Gaia, 2MASS and ALLWISE.
\item {\tt FIELD\_expmap.fits.gz} --  The ``exposure" map per band.
\end{enumerate}

\noindent
More detailed descriptions of the catalogs can be found in the PHOTRED ``README" file\footnote{\url{ftp://archive.noao.edu/public/hlsp/smash/dr1/photred/README}}
on the ftp site (see below).

\subsection{Photometric Precision}
\label{subsec:photprecision}

The photometric precision of the final SMASH catalogs can be estimated by calculating the scatter in multiple independent measurements of the same object using bright stars.
We measured the minimum of median-binned (0.2 mag bins) photometric scatter values
of bright stars for 126 deep and fully-calibrated SMASH fields in each band (see Figure \ref{fig_photscatter}).
The distributions indicate a precision of roughly 1.0\% ($u$), 0.7\% ($g$), 0.5\% ($r$), 0.8\% ($i$), and 0.5\% ($z$) in the SMASH photometry.

\subsection{Photometric Accuracy}
\label{subsec:photaccuracy}

To evaluate the accuracy of the photometric calibration, we use the overlap of fields in the LMC/SMC main-body fields that are independently
calibrated.  Using the scatter in their distributions of mean magnitude offsets (and accounting for the $\sqrt2$ because there are contributions from both fields)
we obtain rough calibration accuracies of 1.3\% ($u$), 1.3\% ($g$), 1.0\% ($r$), 1.2\% ($i$), and 1.3\% ($z$).  The low scatter in the SMASH-Gaia color-color
relations (especially for the redder bands) also attest to the high quality of the SMASH calibration.

\subsection{Photometric Depth}
\label{subsec:photdepth}

The median 5$\sigma$ point source depths in $ugriz$ bands are (23.9, 24.8, 24.5, 24.2, 23.5) mag, respectively, which is $\sim$2 mags deeper than SDSS
and $\sim$1.4 mags deeper than Pan-STARRS1.



\subsection{Astrometric Performance}
\label{subsec:astrometric}

The astrometric precision of the individual measurements of bright stars is $\sim$20 mas.   The precision of the average coordinates of objects (each
having $\sim$10--30 measurements) is $\sim$15 mas and is limited by the systematics in the higher-order WCS distortion terms.  The astrometric
accuracy is $\sim$2 mas per coordinate with respect to the Gaia reference frame.

\begin{figure*}
\begin{center}
$\begin{array}{ccc}
\includegraphics[width=0.40\hsize,angle=0]{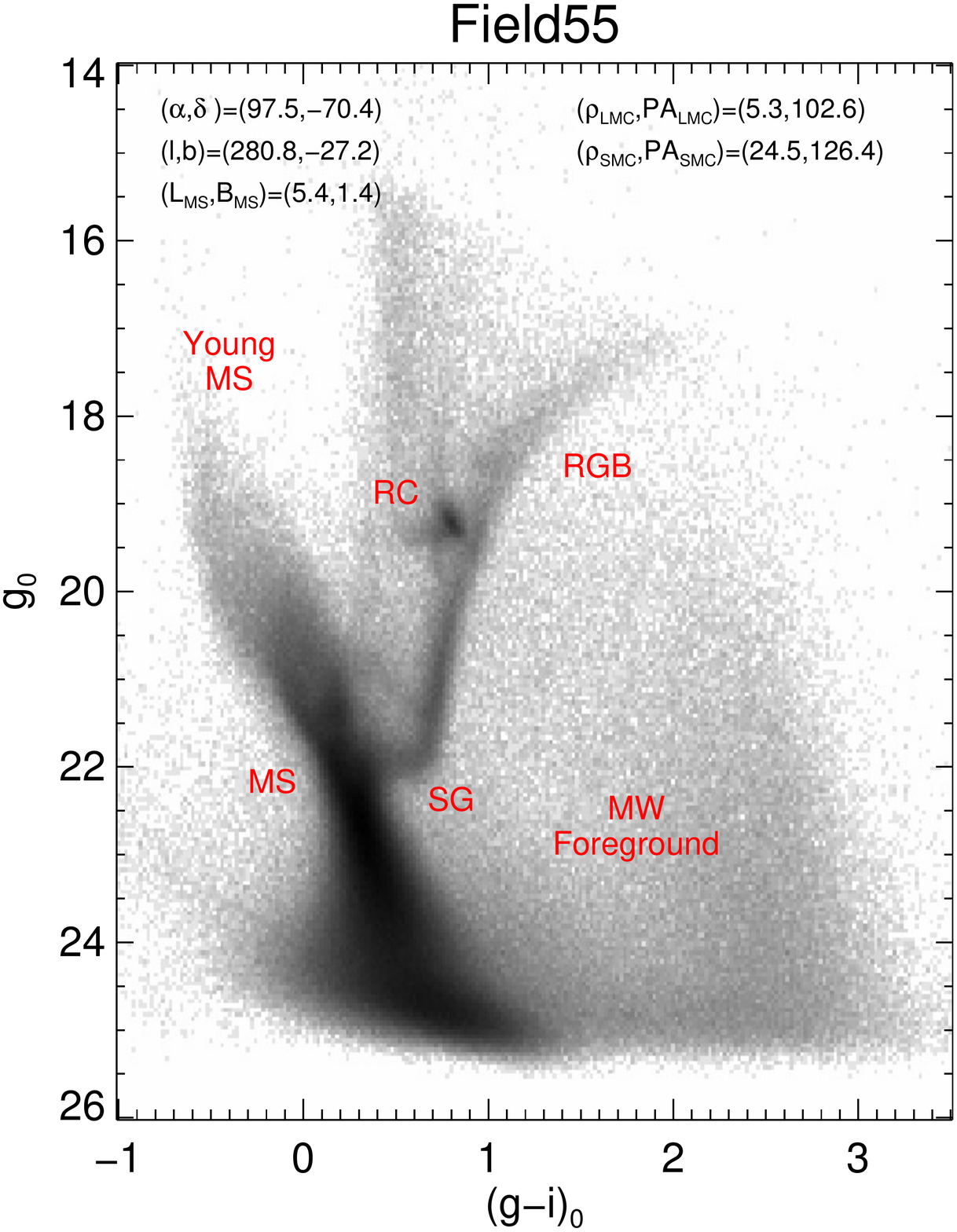}
\includegraphics[width=0.40\hsize,angle=0]{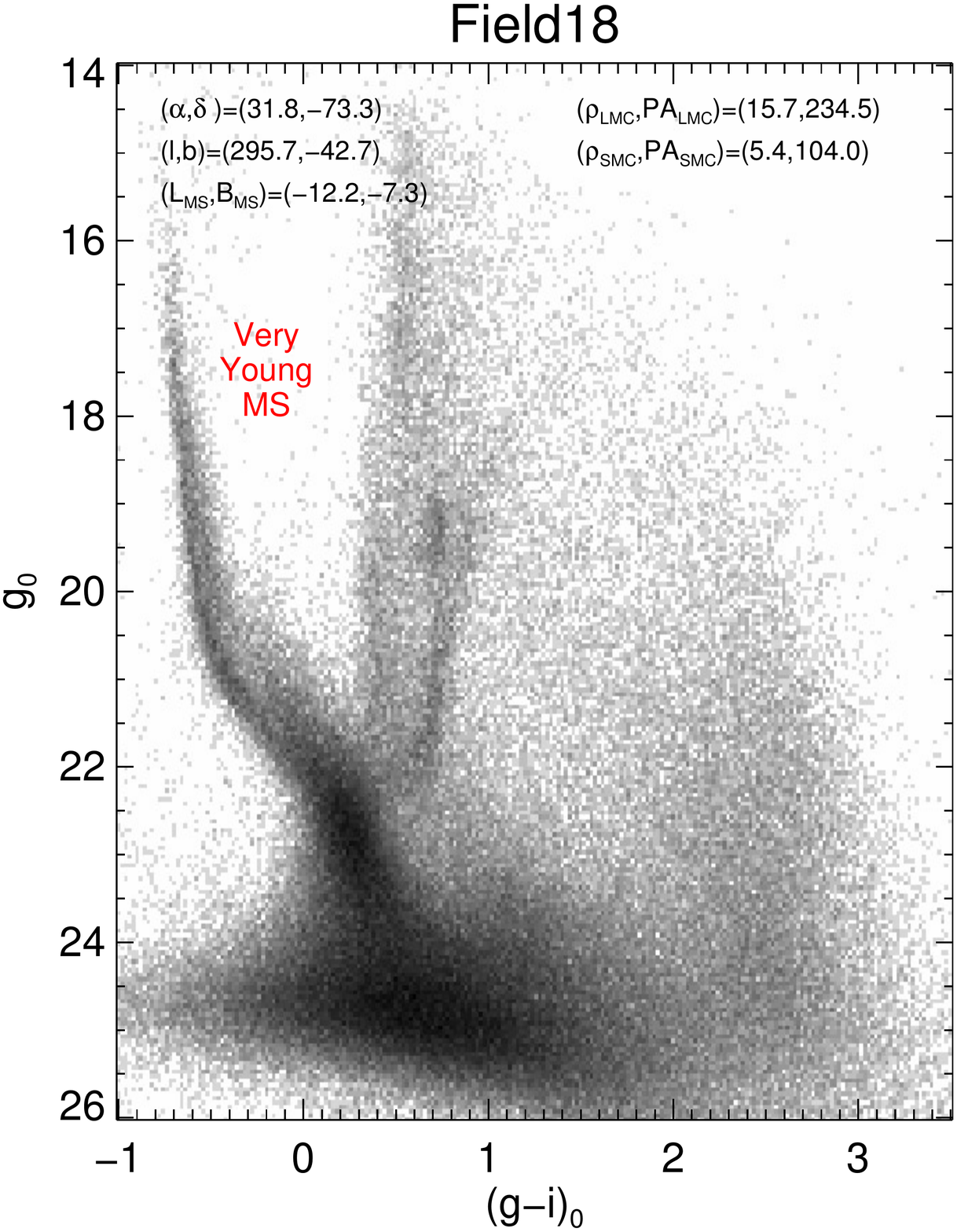} \\
\includegraphics[width=0.40\hsize,angle=0]{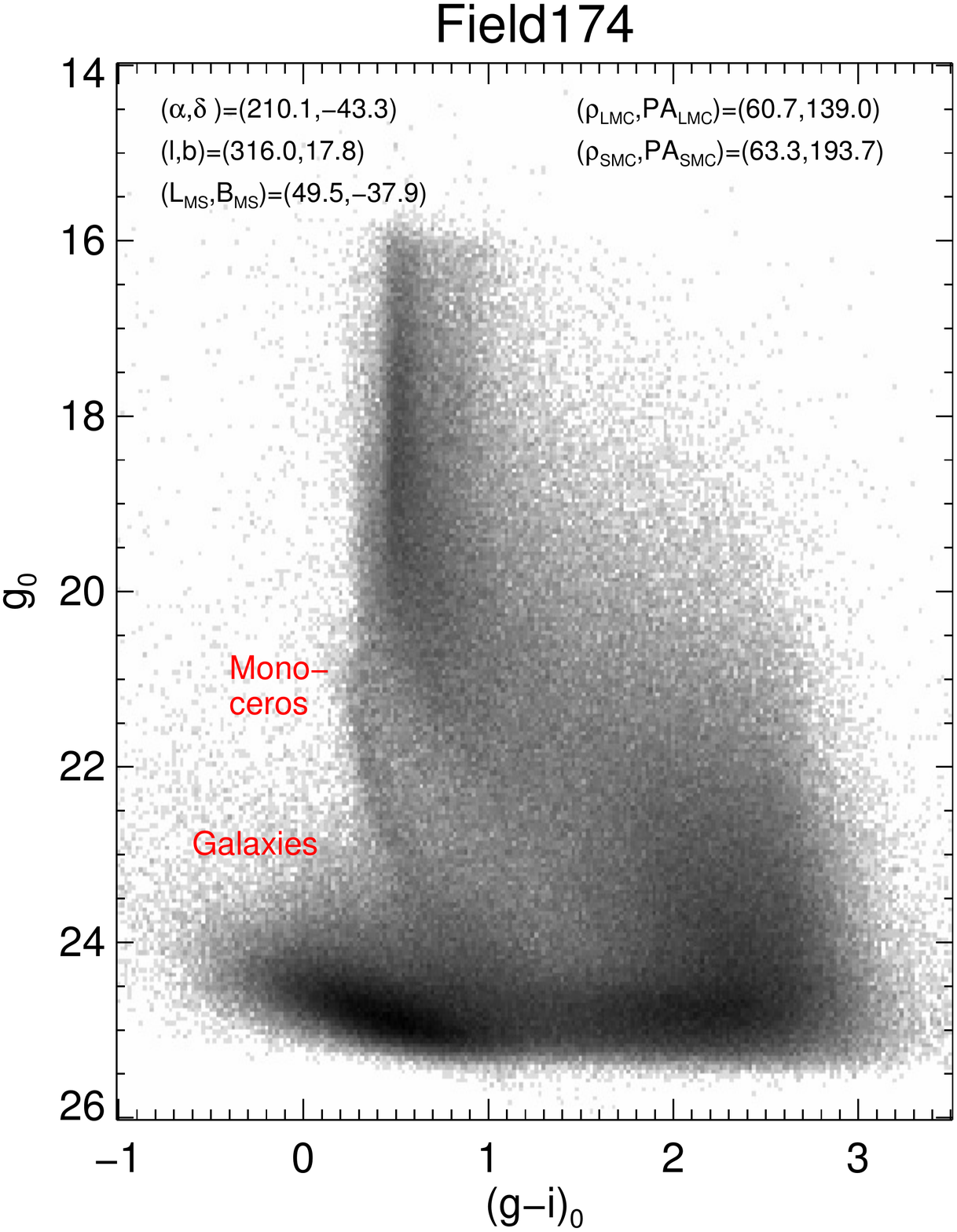}
\includegraphics[width=0.40\hsize,angle=0]{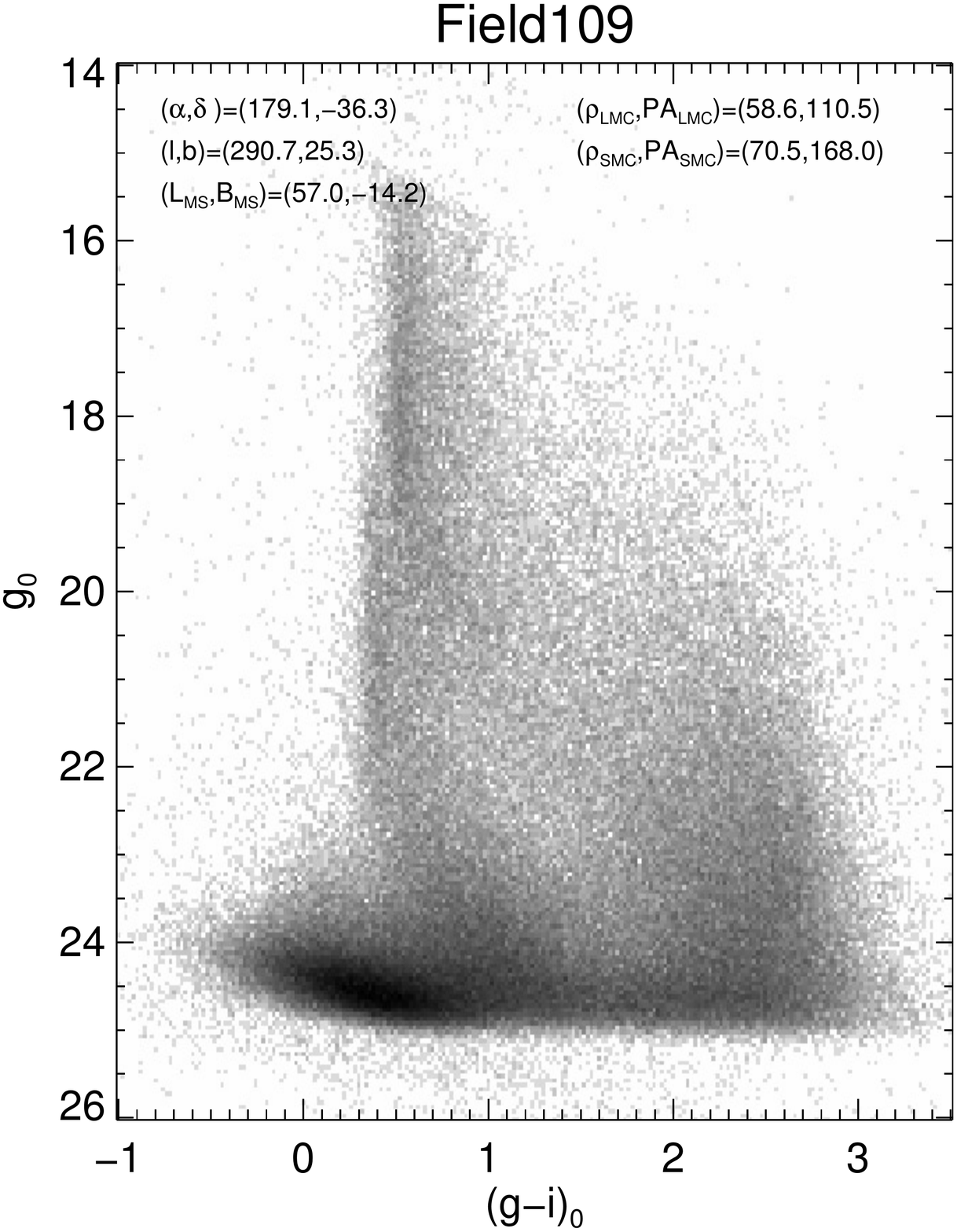}
\end{array}$
\end{center}
\caption{Some example SMASH Hess diagrams illustrating the depth and high-quality of the SMASH photometry and diversity of the stellar populations probed.
Field55 is on the eastern side of the LMC (R=5.3\degr); Field18 is in the Magellanic Bridge region near the SMC; Field174 is in the Leading Arm region but close to the
Milky Way midplane ($b$=17.8\degr); Field109 is also in the Leading Arm region but at higher Galactic latitude ($b$=25.3\degr). Equatorial, Galactic and Magellanic
Stream coordinates are given in the upper-left corner of each panel, while the radius and position angle with respect to the LMC and SMC are given in the upper-right.}
\label{fig_cmds}
\end{figure*}

\section{First Public Data Release}
\label{sec:dr1}


The first SMASH public data release contains $\sim$700 million measurements of $\sim$100 million objects in 61 deep and fully-calibrated fields sampling the $\sim$2400 deg$^2$
region of the SMASH survey (blue hexagons in Figure \ref{fig_dr1map}).  The rest of the data will be included in our second data release in 2018.
The main data access is through a prototype version of the NOAO Data Lab\footnote{\url{http://datalab.noao.edu/}}. 
Access and exploration tools include a custom Data Discovery tool, database access to the catalog (via direct query or TAP service), an image cutout service, and a Jupyter
notebook server with example notebooks for exploratory analysis.  The data release page also gives extensive documentation on the SMASH survey, the observing strategy,
data reduction and calibration, as well as information on the individual data products.

Images, intermediate data products, and final catalogs (in FITS binary formats) are also available through the NOAO High Level Data Products FTP
site\footnote{\url{ftp://archive.noao.edu/public/hlsp/smash/dr1/}}.  The raw images as well as the CP-reduced InstCal, Resampled and single-band Stacked images are
available in {\tt raw/}, {\tt instcal/}, {\tt resampled/}, and {\tt stacked/} directories, respectively (and grouped in nightly subdirectories).  Each subdirectory has a {\tt README}
file that gives information about each FITS image file (e.g., exposure number, time stamp, filter, exposure time, field).  The PHOTRED-ready FITS files and other associated
files (PSF, photometry catalogs, logs, etc.) as well as the multi-band stacks are available in the {\tt photred/} directory.  The final binary FITS catalogs (as described in Section \ref{subsec:catalogdescription})
are in the {\tt catalogs/} directory.  Finally, there are seven tables in the database that were populated using the FITS catalogs (but somewhat modified): field, exposure, chip,
source, object, and xmatch.  The ``field" table includes summary information for each field.  A detailed description of the database schema (tables and columns) is given on
the SMASH data release website.




\begin{figure*}[t]
\includegraphics[width=1.0\hsize,angle=0]{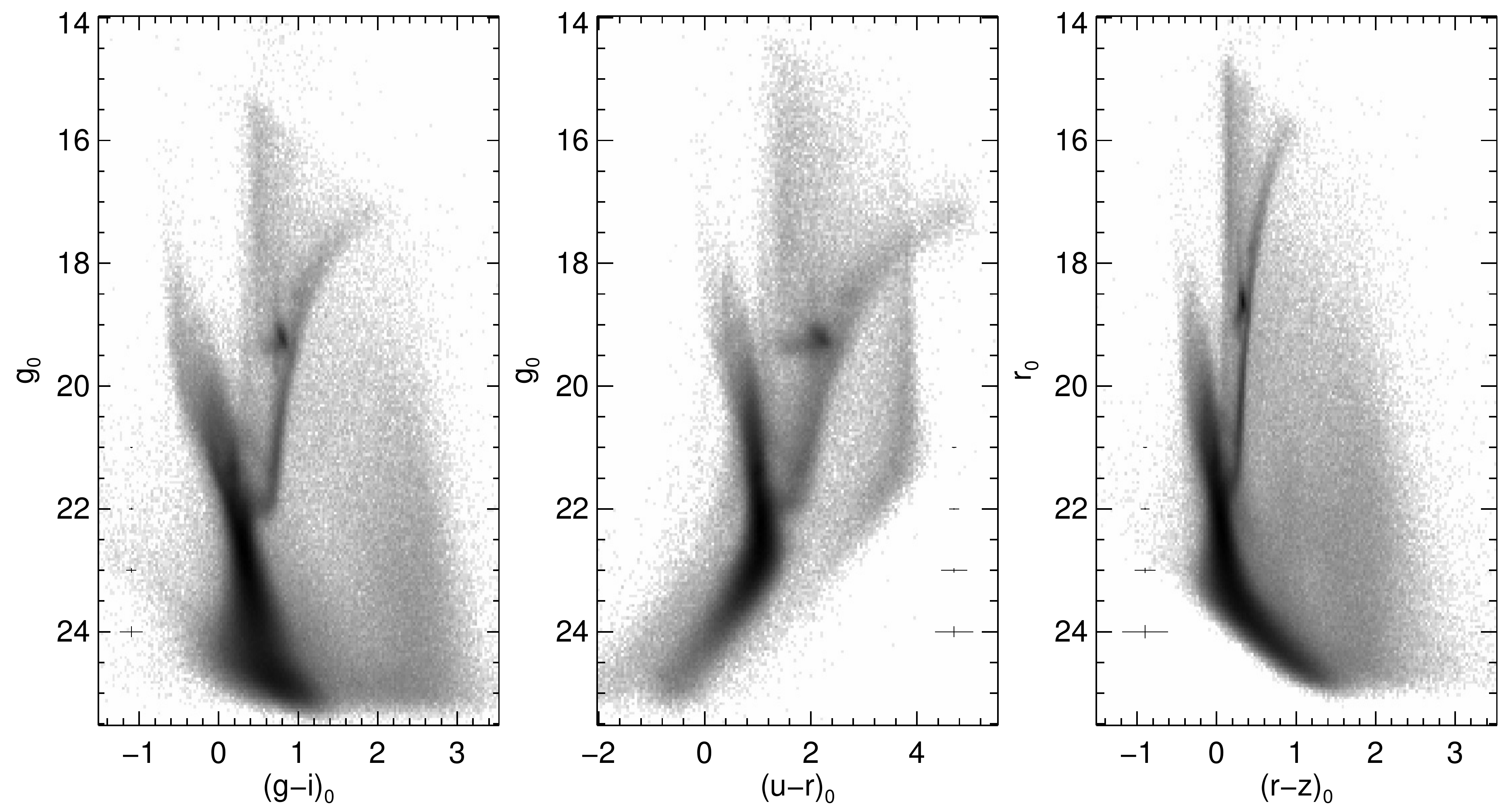}
\caption{Multiple CMDs for Field55 using all $ugriz$ bands.  (Left) $g_0$ vs.\ $(g-i)_0$, (middle) $g_0$ vs.\ $(u-r)_0$, and (right) $r_0$ vs.\ $(r-z)_0$.
Error bars in magnitude and color are shown at the 21st, 22nd, 23rd, and 24th magnitude levels.
}
\label{fig_multicmd}
\end{figure*}

\section{Results}
\label{sec:results}

SMASH is a deep, multi-band photometric survey of the Magellanic Clouds with the goal of mapping the stellar features of these two nearby galaxies to very
low surface brightness and thereby providing better understanding of their joint formation and evolution.  The data in the central regions of the Clouds will also
be used to obtain spatially-resolved star formation histories of these galaxies to old ages.  The data, spanning a region of 2400 deg$^2$, have been
processed and calibrated to high fidelity with almost four billion measurements of 420 million objects in 197 fields.  The data for 61 of these fields and 100
million objects are in the first public data release through the NOAO Data Lab.  Figure \ref{fig_cmds} shows some example Hess diagrams of a number of
our SMASH fields, which indicate the depth ($\sim$2 mag below the oldest main-sequence turnoff in the LMC) and high quality of the SMASH photometry.
In addition, Figure \ref{fig_multicmd} shows multiple CMDs using all $ugriz$ bands for Field55.

The SMASH data have already produced some exciting results.  In \citet{Martin15}, we presented the discovery of a compact and faint Milky Way satellite, Hydra II (in Field169),
with morphological and stellar population properties consistent with being a dwarf galaxy (also see \citealt{Kirby15}).  Interestingly, comparison with simulations suggests that,
given Hydra II's position in the sky and distance of $\sim$140 kpc (from blue horizontal branch stars), it could be associated with the Leading Arm of the Magellanic Stream.
Proper motion information, however,  is needed to confirm this possibility.  We obtained follow-up time-series data on Hydra II to study its variable stars.
This work yielded one RR Lyrae star in Hydra II that gave a slightly larger distance of 151$\pm$8 kpc, as well as the discovery of dozens of short period
variables in the field \citep{Vivas16}.

Further sensitive searching for overdensities in the SMASH data yielded the discovery of a compact and very faint ($M_V$ = $-$1.0) stellar system
(designated SMASH 1) $\sim$11\dgr away from the LMC \citep{Martin16}.
SMASH 1 is consistent with being an old globular cluster in the LMC periphery likely associated with the LMC disk and potentially on the verge of being tidally disrupted.

One of the on-going SMASH projects is to map out the extended stellar populations of the LMC.  An analysis of the Hess diagrams indicates that LMC stellar populations can
be detected in SMASH data out to 21.1\dgr from the LMC center\footnote{We use ($\alpha$,$\delta$)=(05:27:36, $-$69:52:00) for the LMC center and ($\alpha$,$\delta$)=(00:52:44, $-$72:49:42) for the SMC center.}, or $\sim$18.4 kpc, and to surface brightness levels of $\sim$33.3 mag arcsec$^{-2}$ (D. Nidever et al. 2017, in preparation).

One of the main goals of SMASH is to use the data in the central LMC/SMC fields to derive spatially-resolved star formation histories.  The Hess diagram of Field55 in Figure \ref{fig_cmds}
(upper left) is an example of the wealth of information in the data.  This field, and other nearby ones show two subgiant branches, which indicates two periods of peak star formation.
This was previously only seen in star formation rate diagrams from detailed star formation history modeling \citep{HZ09,Meschin14}, but now is visually clear just in the Hess diagrams.
Full star formation history modeling still awaits computationally intensive artificial star tests for the SMASH data, which will be a focus of on-going SMASH processing efforts in the near future.

The deep and multi-band data in the main bodies of the Magellanic Clouds are also very useful for detecting faint star clusters.  We are in the process of developing a citizen science project
(led by L.C.J.)  based on the SMASH data under the Zooniverse platform\footnote{\url{https://www.zooniverse.org}}  which currently has roughly one million users and hosts many citizen science
projects in multiple scientific disciplines.  The project will be called ``The Magellanic Project" and will be similar to the ``The Andromeda Project" of $HST$ images of M31.  The citizen scientists
will inspect our deep co-add $ugriz$ images and visually identify (a) star clusters (open and globular), (b) galaxies behind the LMC/SMC main bodies, (c) and potential
new dwarf galaxies of the MW or the MCs.  The website is projected to be launched in early 2018.

The SMASH data are also very useful for studying structures in the MW halo that are unrelated to the MCs.  The Hess diagram of Field174 in Figure \ref{fig_cmds} (lower left panel), 
not far above the MW mid-plane ($b$=17.8\degr), shows a prominent stellar population at a distance of $\sim$10--20 kpc
(thin sequence with ($g$$-$$i$)$_0$$\sim$0.5 and 21.0$\lesssim$$g_0$$\lesssim$23.5).
Many other fields at low Galactic latitude show similar stellar populations that are very likely associated with the Monoceros ``ring" \citep[e.g.,][]{Slater14}.
There are on-going SMASH projects to study these and similar MW halo structures in the SMASH data.


The SMASH survey has the potential to revolutionize our understanding of the stellar populations inside and in the very outskirts of two canonical examples of dwarf galaxies,
the SMC and LMC.

\acknowledgments

DLN was supported by a McLaughlin Fellowship while at the University of Michigan. YC acknowledges support from NSF grant AST 1655677.
BCC acknowledges the support of the Australian Research Council through Discovery project DP150100862.
EFB acknowledges support from NSF grants AST 1008342 and 1655677.
EWO was partially supported by NSF grant AST 1313006.
TdB acknowledges financial support from the ERC under Grant Agreement n.\ 308024.
M-RC acknowledges support by the German Academic Exchange Service (DAAD), from the UK's Science and Technology Facility Council [grant number ST/M001008/1],
and from the the European Research Council (ERC) under the European Union's Horizon 2020 research and innovation programme (grant agreement No 682115).
SJ is supported by the Netherlands Organization for Scientific Research (NWO) Veni grant 639.041.131.
SRM acknowledges partial support from NSF grant AST 1312863.
DM-D acknowledges support by Sonderforschungsbereich (SFB) 881 ``The Milky Way System" of the
German Research Foundation (DFB), subproject A2. RRM acknowledges partial support from CONICYT Anillo project ACT-1122 and project BASAL PFB-$06$. GSS is supported by grants from NASA.
Based on observations at Cerro Tololo Inter-American Observatory, National Optical Astronomy Observatory (NOAO Prop. ID: 2013A-0411 and 2013B-0440; PI: Nidever), which is operated by the Association of Universities for Research in Astronomy (AURA) under a cooperative agreement with the National Science Foundation.
IRAF is distributed by the National Optical Astronomy Observatory, which is operated by the Association of Universities for Research in Astronomy (AURA) under a cooperative agreement with the National Science Foundation.
This project used data obtained with the Dark Energy Camera (DECam), which was constructed by the Dark Energy Survey (DES) collaboration. Funding for the DES Projects has been provided by the U.S. Department of Energy, the U.S. National Science Foundation, the Ministry of Science and Education of Spain, the Science and Technology Facilities Council of the United Kingdom, the Higher Education Funding Council for England, the National Center for Supercomputing Applications at the University of Illinois at Urbana-Champaign, the Kavli Institute of Cosmological Physics at the University of Chicago, Center for Cosmology and Astro-Particle Physics at the Ohio State University, the Mitchell Institute for Fundamental Physics and Astronomy at Texas A\&M University, Financiadora de Estudos e Projetos, Funda\c{c}\~ao Carlos Chagas Filho de Amparo, Financiadora de Estudos e Projetos, Funda\c{c}\~ao Carlos Chagas Filho de Amparo \`a Pesquisa do Estado do Rio de Janeiro, Conselho Nacional de Desenvolvimento Cient�fico e Tecnol\'ogico and the Minist\'erio da Ci\^encia, Tecnologia e Inova\c{c}\~ao, the Deutsche Forschungsgemeinschaft and the Collaborating Institutions in the Dark Energy Survey. The Collaborating Institutions are Argonne National Laboratory, the University of California at Santa Cruz, the University of Cambridge, Centro de Investigaciones En\'ergeticas, Medioambientales y Tecnol\'ogicas-Madrid, the University of Chicago, University College London, the DES-Brazil Consortium, the University of Edinburgh, the Eidgen\"ossische Technische Hochschule (ETH) Z\"urich, Fermi National Accelerator Laboratory, the University of Illinois at Urbana-Champaign, the Institut de Ci\`encies de l'Espai (IEEC/CSIC), the Institut de F\'isica d'Altes Energies, Lawrence Berkeley National Laboratory, the Ludwig-Maximilians Universit\"at M\"unchen and the associated Excellence Cluster Universe, the University of Michigan, the National Optical Astronomy Observatory, the University of Nottingham, the Ohio State University, the University of Pennsylvania, the University of Portsmouth, SLAC National Accelerator Laboratory, Stanford University, the University of Sussex, and Texas A\&M University.
This work has made use of data from the European Space Agency (ESA) mission {\it Gaia} (\url{https://www.cosmos.esa.int/gaia}), processed by
the {\it Gaia} Data Processing and Analysis Consortium (DPAC, \url{https://www.cosmos.esa.int/web/gaia/dpac/consortium}). Funding
for the DPAC has been provided by national institutions, in particular the institutions participating in the {\it Gaia} Multilateral Agreement.
This publication makes use of data products from the Two Micron All Sky Survey, which is a joint project of the University of Massachusetts and the Infrared Processing and Analysis Center/California Institute of Technology, funded by the National Aeronautics and Space Administration and the National Science Foundation.

\end{document}